\documentclass[a4paper,11pt]{article}
\pdfoutput=1 

\usepackage{jheppub} 

\usepackage[T1]{fontenc} 

\usepackage{amsmath,amssymb,amsfonts}
\usepackage{graphicx}
\usepackage{hyperref}
\usepackage{bbm}
\usepackage{mathdots}
\usepackage{color}
\usepackage{ytableau}
\ytableausetup{aligntableaux=center,smalltableaux}
\usepackage{tabularx}
\usepackage{here}
\usepackage{comment}
\usepackage{makecell}

\usepackage[nameinlink]{cleveref}
\crefname{section}{}{}
\creflabelformat{section}{#2Sec.~#1#3}
\crefname{equation}{}{}
\creflabelformat{equation}{#2Eq.~(#1#3)}

\def\cW{\mathcal{W}}

\def\bZ{\mathbb{Z}}

\def\tr{\mathrm{tr}}

\def\det{\mathrm{det}}
\def\Pf{\mathrm{Pf}}

\def\rSpin{{\mathrm{Spin}}}

\usepackage{cellspace}
\newcommand{\myspace}[1]{
    \setlength{\cellspacetoplimit}{#1mm}
    \setlength{\cellspacebottomlimit}{#1mm}
}

\title{\boldmath AMSB in Truly Confining Gauge Theories}

\author[a]{Riku Ishikawa,}
\author[a,b,c,1]{Hitoshi Murayama,\note{Hamamatsu Professor}}
\author[a]{Shota Saito}

\affiliation[a]{Kavli Institute for the Physics and Mathematics of the Universe (WPI), UTokyo Institutes for Advanced Study, University of Tokyo, Kashiwa 277-8583, Japan}
\affiliation[b]{Department of Physics, University of California, Berkeley, California 94720, USA}
\affiliation[c]{Theoretical Physics Group, Lawrence Berkeley National Laboratory, Berkeley, California 94720, USA}

\emailAdd{riku.ishikawa@ipmu.jp}
\emailAdd{hitoshi@berkeley.edu}
\emailAdd{shota.saito@ipmu.jp}

\abstract{
We study deformation of supersymmetric $t$-confining (truly confininig) gauge theories with small anomaly mediation of supersymmetry breaking. We identify breaking pattern of global symmetries in the theories. These results can be compared in priciple to lattice simulation of non-supersymmetric theories (except for one of the cases where the theory is pseudoreal and chiral). 
}

\begin{document} 
\maketitle
\flushbottom

\section{Introduction}
\label{sec:intro}

Confinement in gauge theories is a genuinely non-perturbative phenomenon and is not easy to understand by analytic means. The word confinement is used to describe the lack of isolated quarks in QCD even though it is strictly not a confinement but rather screening by light quarks in the fundamental representation. The most precise definition of confinement is vanishing of the Polyakov loop $P = \langle {\rm Tr} T e^{i\oint A d t}\rangle$ along the Euclidean time direction. $P$ is charged under the one-form symmetry $P \rightarrow P Z$ where $Z$ is an element of the center of the gauge group $G$ \cite{Gaiotto:2014kfa}. When $P=0$, which is physically interpreted as an infinite energy of an isolated quark, the one-form symmetry is unbroken. When a quark can be isolated with a finite energy $E_q$, $P \simeq e^{-\beta E_q} \neq 0$ where $\beta$ is the Euclidean time interval, the one-form symmetry is spontaneously broken. Therefore two phases can be clearly distinguished from each other. Note that this definition requires a center element that cannot be screened by light quarks in the theory. In this case, we say the theory is ``truly confined'' or $t$-confined.\footnote{This definition does not preclude the possibility that a theory can have a vanishing Polyakov looop despite the absence of a center, see, {\it e.g.}\/, \cite{Holland:2003jy}. } 

We classified all $t$-confining theories with $N=1$ supersymmetry in \cite{Ishikawa:2025yus} (see Table~\ref{$t$-confinement}). Interestingly, many of them have multiple branches either with an Affleck--Dine--Seiberg-like run-away superpotential or a moduli space without a superpotential. The non-trivial anomaly matching including discrete symmetries provide strong evidence for $t$-confinement in these theories. Often one of the discrete symmetries is partially broken by ``magnetic operators'' with an exception of class (C) theories. 

\begin{table}[h]
    \centering
    \myspace{2}
    \begin{tabular}{ScSlSlSl}
        \hline
        & Theory & ``Magnetic'' Operator & Discrete Symmetry \\
        \hline
        (A) & $G$ & $\tr(\cW^\alpha\cW_\alpha)$ & $\bZ_{2h}\to\bZ_2$ \\
        (B) & $\rSpin(k)+(k-4)V^i$ & $\tr(V^{k-4}\cW^\alpha\cW_\alpha)$ & $\bZ_{2k-8}\to\bZ_{k-4}$ \\
        (C) & $\rSpin(k)+(k-3)V^i$ & no operator & no breaking \\
        (D) & $SU(6)+A^{ijk}$ & $\tr(A^2\cW^\alpha\cW_\alpha)$ & $\bZ_6\to\bZ_2$ \\
        (E) & $Sp(k)+A^{ij}$ & $\tr(A^{k-1}\cW^\alpha\cW_\alpha)$ & $\bZ_{2k-2}\to\bZ_{k-1}$ \\
        (F) & $SU(2k)+A^{ij}+\tilde{A}_{ij}$ & $\tr\{(A\tilde{A})^{k-1}\cW^\alpha\cW_\alpha\}$ & $\bZ_{4k-4}\to\bZ_{2k-2}$ \\
        (G) & $\rSpin(12)+2S$ & $\tr(S^8\cW^\alpha\cW_\alpha)$ & $\bZ_{16}\to\bZ_{8}$ \\
        \hline
    \end{tabular}
    \caption{Summary of dynamics in the confinement phase of $t$-confining theories \cite{Ishikawa:2025yus}}
    \label{$t$-confinement}
\end{table}

Having identified $t$-confining theories with supersymmetry, the next natural question is their non-supersymmetric counter parts. In particular, the pattern of symmetry breaking is an interesting question. Unfortunately we do not have analytic tools to study non-supersymmetric gauge theories non-perturbatively even though lattice simulations may be possible for some of them.

What we study in this paper is the deformation of supersymmetric $t$-confining theories by anomaly mediation of supersymmetry breaking (AMSB) \cite{Randall:1998uk,Giudice:1998xp}. When the scale of AMSB $m$ is much smaller than the dynamical scale of the gauge theory $\Lambda$, the UV-insensitive nature of AMSB allows for exact non-perturbative solutions \cite{Murayama:2021xfj}. Using this methodology, we can study non-supersymmetric $t$-confining gauge theories and identify their symmetry breaking patterns. At least for QCD-like theories, there is an ample evidence that the near SUSY limit $m \ll \Lambda$ and non-SUSY limit $m \gg \Lambda$ are continuously connected as a cross over without a phase transition \cite{Kondo:2025njf}.

One tricky case is the branch without a superpotential. Naively AMSB vanishes on the moduli space with free composite chiral superfields because of an emergent conformal invariance. However the K\"ahler potential does violate the conformal invariance and AMSB does induce a non-zero potential. Unfortunately K\"ahler potential is not constrained by holomorphy and we cannot compute it from the first principle. Nonetheless the K\"ahler potential can produce a minimum away from the origin as shown in the App.~\ref{sec:Kahler}. In most examples, it is highly implausible that there are massless composite fermions once the constituent scalar acquires mass in the UV due to AMSB. If so, anomalies cannot be matched by massless fermions and hence global symmetries must be broken. This is indeed what we find in most of the examples, while the cases (D) and (G) (see Table~\ref{$t$-confinement}) remain ambiguous. Yet this branch gives a shallower minimum compared to the branches with non-zero superpotential. Therefore the global minimum is always on the branch with a superpotential. 

We point out that three physical cases need to be considered, depending on the SUSY-breaking scale $m$.
The SUSY vacuum corresponds to $m=0$, the vacuum perturbed by AMSB corresponds to $m\ll\Lambda$, and the non-SUSY vacuum corresponds to $m\gg\Lambda$.
In the multibranch structure typically found in $t$-confining theories, different branches are expected to have different minima.
Given that the branches appear due to the gaugino condensation in unbroken gauge groups, we expect them to be separated by domain walls.
Such domain walls have 3D Topological Quantum Field Theories (TQFTs) that include Chern--Simons terms on them \cite{Acharya:2001dz,Delmastro:2020dkz}.
Presumably tunneling is possible across the domain walls so that the theory settles on the the global minimum among all branches.
Therefore, there is expected to be a phase transition between the vacuum at $m=0$ and the vacuum for $m\ll\Lambda$.
However, the vacua with $m\ll\Lambda$ and $m\gg\Lambda$ are still expected to be continuously connected with no phase transition \cite{Kondo:2025njf}.

The organization of this paper is as follows.
In Sec.~\ref{sec:review}, we review the criteria of $t$-confinement in terms of unscreened center symmetry and Wilson line order parameters.
In addition, we introduce AMSB and show some basis examples of its application to SQCD.
We discuss pure Yang-Mills theory and $\rSpin(k)$ gauge theories with vectors \cite{Csaki:2021jax,Csaki:2021xuc} perturbed by AMSB, where both of them are $t$-confining theories.
From Sec.~\ref{sec:SU(6)+A} to Sec.~\ref{sec:Spin(12)+2s}, we discuss AMSB aspects of $SU(6)$ with a rank three antisymmetric tensor, $Sp(k)$ with an antisymmetric tensor, $SU(2k)$ with an antisymmetric tensor and its congugate, and $\rSpin(12)$ with two spinors, respectively.
We derive effective potentials, determine vacuum structures, and identify symmetry breaking patterns.
Above four classes of $t$-confining theories possess multiple branches, and these features differ from one branch to another unless they are related by discrete symmetry transformations.
In particular, we show that SUSY and non-SUSY branches are distinct and therefore exhibit different patterns of symmetry breaking.
These four sections can be read independently and in parallel. In App.~\ref{sec:Kahler}, we show how braches without a superpotential can still find potential due to AMSB because of non-canonical K\"ahler potential which can be at the origin respecting global symmetries or away from the origin breaking a part of the global symmetries.

\section{Truly confinement and Anomaly mediation}
\label{sec:review}

In this section, we give an brief introduction to truly confining ($t$-confining) supersymmetric gauge theories \cite{Ishikawa:2025yus}, and anomaly mediated supersymmetry breaking (AMSB) \cite{Murayama:2021xfj}.
In addition, as examples of applying AMSB to $t$-confining theories, we introduce three theories, that is, pure Yang-Mills, $\rSpin(k)$ with $k-4$ vectors, and $\rSpin(k)$ with $k-3$ vectors \cite{Csaki:2021jax,Csaki:2021xuc}.

\paragraph{Truly confinement}
Here, we review $t$-confining SUSY gauge theories \cite{Ishikawa:2025yus}.
The main point is that $t$-confinement is not meant to describe confinement in the sense of the absence of colored asymptotic states.
Rather, it is a notion tied to the existence of Wilson line order parameters with a nontrivial unscreened center symmetry that cannot be screened by dynamical matter.
In this sense, $t$-confining theories are those in which a genuine distinction remains between the confining and Higgs regimes.

There are three conditions for $t$-confining theories.
First, as mentioned above, the center symmetry must not be completely screened \cite{Aharony:2013hda}.
This is the condition of the existence of a nontrivial electric one-form symmetry.
Let $G$ be a gauge group of a theory.
If the matter fields have center charges $q_i$, the unscreened condition is given by
\begin{equation}
    Z(G)/\gcd(q_1,q_2,\cdots,q_m)\neq 1.
\end{equation}
The minimal Wilson line operator is charged under this remaining center symmetry, and its large-loop expectation value distinguishes different phases of the gauge theory exhibiting an area law in a confining phase \cite{wilson1974confinement,PhysRevD.19.3682,BANKS1979349}.
Second, an index condition is needed to exclude unpromising theories and to reduce the candidate theories to a finite class.
Denote by $I_i$ the Dynkin index of the $i$-th matter representation and by $I_G$ the Dynkin index of the adjoint representation. The condition is given by
\begin{equation}
    \sum_i I_i < I_G+2.
\end{equation}
This condition is empirical rather than derived purely from symmetry.
Its role is to isolate theories that can plausibly have a confining phase from the much larger set of asymptotically free SUSY gauge theories.
When the inequality is reversed, the theory is typically expected to flow instead to an $s$-confining, infrared-free, or non-Abelian Coulomb phase, rather than to a $t$-confining phase. 
Third, we impose ’t Hooft anomaly matching conditions \cite{hooft1980naturalness}, including discrete symmetries \cite{csaki1998discrete}.
This condition must always be satisfied by the theory.
Here, in particular, it is used to verify whether the gauge-invariant operators properly describe the IR.
For $t$-confining theories, this requirement is especially important because discrete symmetries often diagnose the condensation of magnetic objects.
If an anomaly involving a discrete symmetry fails to match in a naive IR description, the interpretation is that the discrete symmetry is spontaneously broken.
This breaking is often caused by the vacuum expectation value of a “magnetic” operator. 

The $t$-confining theories with a simple Lie group and no tree-level superpotential are summarized in Table~\ref{$t$-confinement}.
A common theme in the classified theories is the condensation of magnetic operators.
In pure $\mathcal N=1$ Yang-Mills theory, the relevant condensate is the gaugino bilinear $\tr(\mathcal{W}^\alpha\mathcal{W}_\alpha)$ which breaks the $R$-symmetry as $\mathbb Z_{2h}\to\mathbb Z_2$ where $h$ is the dual Coxeter number of the gauge group.
This is the prototype for the more general pattern.
In $t$-confining theories, a composite operator like $\tr\{(\text{matter})\mathcal{W}^\alpha\mathcal{W}_\alpha\}$ condenses and breaks a discrete symmetry.
These condensates are interpreted as magnetic order parameters for the confining dynamics.
In most cases, they are directly visible in the low-energy analysis, while in the $\rSpin(k)$ theory with $k-3$ vectors the relevant magnetic condensation is instead understood by embedding the theory in an upstream theory with an additional flavor and decoupling it.
Another important feature is the appearance of multiple branches.
Except for pure Yang-Mills theory and $\rSpin(k)$ theory with $k-3$ vectors, the $t$-confining theories exhibit a branch with vanishing superpotential $W=0$.
The existence of such branches is notable because they arise from choices of signs in dynamically generated superpotentials and are tied to the multi-branch structure of the moduli space.
This feature is important for AMSB deformations, since anomaly mediated soft terms can lift different branches in different ways.

For the purposes of the present paper, the essential lesson is that $t$-confining theories provide a controlled class of SUSY gauge theories in which confinement is tied to an unscreened center symmetry and to the condensation of magnetic objects.
This makes them especially suitable for studying AMSB.
AMSB can be applied in the regime $m\ll \Lambda$ where the exact IR description remains reliable.
One can then ask how the AMSB induced scalar potential lifts the moduli space, selects among branches, and affects the magnetic condensates that diagnose confinement.
This is precisely the perspective we will develop in the following sections.

Finally, we comment on the global structure of the gauge group.
In general, the unscreened condition does not uniquely determine the fundamental group of the gauge group.
There may remain the possibility of quotienting the universal covering group by a subgroup of its center.
In order to identify it, let us consider a gauge group $\tilde{G}$ with a center $\bZ_n$ and its matter in a $N$-ality representation $R$.
When we consider a quotient group $\tilde{G}/\bZ_p$, $N/p$ must be an integer for $R$ to be a representation of this group.
The electric one-form symmetries are $\bZ_{\mathrm{gcd}(n,N)}$ and $\bZ_{\mathrm{gcd}(n,N)/p}$ respectively.
Therefore, a $t$-confining theory with a prime $N$, the gauge group should be simply-connected.
Note that theories with gauge group $\rSpin(4k)$, whose center is not cyclic, can be analyzed in a similar manner.
As a result, all $t$-confining theories with matter fields have simply-connected gauge groups.
For pure Yang-Mills theory, however, there remains the possibility of quotienting the gauge group by a subgroup of its center.

\paragraph{Anomaly mediation}
AMSB is a useful way of introducing soft SUSY breaking into a SQCD \cite{Seiberg:1994bz,Seiberg:1994pq,Intriligator:1994jr}.
The central idea is that SUSY breaking is transmitted to the visible sector through the super Weyl anomaly, generating loop-level SUSY breaking effects on tri-linear couplings, scalar masses, and gaugino masses \cite{Randall:1998uk,Giudice:1998xp}.
AMSB can be derived within the framework of supergravity theory \cite{Pomarol:1999ie,Jack:1999aj,Sanford:2010hc,DEramo:2012vvz}.
In particular, AMSB has the remarkable property of being UV-insensitive in the sence that the AMSB soft terms are determined by the RG functions of the effective theory at each energy scale.
Therefore AMSB can be applied to the theory at any energy scale using an appropriate description in that scale.
This feature makes AMSB highly predictive and, more importantly, allows for an exact description of the strongly coupled regime.

A convenient way to describe AMSB is to introduce the compensator $\Phi$. In a flat-space effective description, SUSY breaking is encoded by its expectation value
\begin{equation}
    \Phi = 1 + m\theta^2 ,
\end{equation}
where $m$ is a SUSY breaking scale. This appears in the SUSY Lagrangian as follows:
\begin{equation}
    \mathcal{L}=\int d^4\theta\Phi\Phi^*K+\int d^2\theta\Phi^3W+\mathrm{h.c.}
\end{equation}
When a theory is exactly conformal, AMSB effects vanish.
When conformal invariance is broken, soft terms are generated.
Then, the AMSB soft terms are given as follows:
\begin{subequations}
\begin{align}
    A_{ijk}(\mu)&=-\dfrac{1}{2}(\gamma_i+\gamma_j+\gamma_k)(\mu)m, \\
    m_i^2(\mu)&=-\dfrac{1}{4}\dot{\gamma}_i(\mu)m^2, \\
    m_\lambda(\mu)&=-\dfrac{\beta(g^2)}{2g^2}(\mu)m,
\end{align}
\end{subequations}
where $\gamma_i=\mu\frac{\partial}{\partial\mu}\ln Z_i(\mu)$, $\dot{\gamma}_i=\mu\frac{\partial}{\partial\mu}\gamma_i$, and $\beta(g^2)=\mu\frac{\partial}{\partial\mu}g^2$.
These formulae show that the AMSB soft terms are determined locally along the renormalization group trajectory. This is the sense in which AMSB is ultraviolet insensitive: once the low-energy effective theory and its renormalization group functions are known, the corresponding soft terms are fixed independently of many details of the ultraviolet completion. This makes AMSB a powerful deformation of SUSY gauge theories whose SUSY dynamics are exactly known.

One of the authors, Murayama, proposed using SQCD perturbed by AMSB as a controlled approximation to QCD-like theories \cite{Murayama:2021xfj,Csaki:2022cyg}.
Research is also underway on applying AMSB to s-confining theories, which represent a concept opposite to that of t-confining theories \cite{Luzio:2022ccn,deLima:2023ebw}.
Such theories are calculable in the limit $m \ll \Lambda$, whereas in the opposite limit, $m \gg \Lambda$, they are expected to approach ordinary non-SUSY gauge theories.
The central physical question is whether these two regimes are continuously connected.
If no phase transition separates them, the analysis in the small-AMSB regime provides exact information about the phase structure of the corresponding non-SUSY theory.
There is a discussion of the possible existence of a phase transition \cite{Dine:2022req}.
This possibility is the main motivation behind using AMSB as a bridge between SUSY and non-SUSY dynamics.
The tree level contribution of AMSB-perturbated SUSY QCD is given by
\begin{equation}
\label{eq:AMSBtree}
    V_{\text{tree}}=\partial_i W\, g^{ij^\ast}\, \partial_{j}^* W^\ast+ m^\ast m\left(\partial_i K\, g^{ij^\ast}\, \partial_{j}^* K- K\right)+ m\left(\partial_i W\, g^{ij^\ast}\, \partial_{j}^* K- 3W\right)+ \text{c.c.}
\end{equation}
where $g^{ij^\ast}$ is the inverse of the K\"{a}hler metric.
\footnote{
    This formula can be derived easily.
    The supersymmetric Lagrangian for a chiral superfield $X^i$ with the Weyl compensator $\Phi=1+\theta^2m$ is written as $\mathcal{L}=\int d^4\theta\bar{\Phi}\Phi K(X^i,\bar{X}^{\bar{i}})+\int d^2\theta\Phi^3W(X^i)+\mathrm{h.c.}$ Expanding the Lagrangian by denoting $X^i$ as $x^i+\theta^2F^i$, we obtain $\mathcal{L}=g_{i\bar{j}}F^i\bar{F}^{\bar{j}}+F^i(W_{,i}+\bar{m}K_{,i})+\bar{F}^{\bar{j}}(\bar{W}_{,\bar{j}}+mK_{,\bar{j}})+|m|^2K+3mW+3\bar{m}\bar{W}$.
    Solving the equation of motion for auxiliary fields and substituting it back, we obtain $\mathcal{L}=-(W_{,i}+\bar{m}K_{,i})g^{i\bar{j}}(\bar{W}_{,\bar{j}}+mK_{,\bar{j}})+|m|^2K+3mW+3\bar{m}\bar{W}$.
    Expanding this expression and flipping the overall sign, we finally obtain Eq.~\eqref{eq:AMSBtree}.
    Here, derivatives are denoted by commas.
}
When the K\"{a}hler potential is canonical, the second term vanishes. Note that in confining theories, the K\"{a}hler potential is often unknown beyond its leading terms. Since the scalar potential depends on the inverse K\"{a}hler metric, higher-order corrections can affect vacuum stability.

\paragraph{AMSB in pure Yang-Mills}
Here we study how the $\mathcal N=1$ Yang-Mills theory based on a simply-connected compact Lie group $G$ behaves when perturbed by AMSB \cite{Veneziano:1982ah,DAVIES1999123}.
We will see that the degeneracy among the multiple vacua present in the supersymmetric theory vanishes.
The AMSB deformation generates a soft mass for the gluino, so that the theory interpolates between pure $\mathcal N=1$ Yang-Mills at small supersymmetry breaking and ordinary pure Yang-Mills after the gluino decouples.
The $\mathcal N=1$ Yang-Mills theory is believed to be gapped, and has $h^\vee_G$ vacua distinguished by the phase of the gluino condensate,
\begin{equation}
    \langle \lambda\lambda\rangle\sim\Lambda^3\exp\left(\dfrac{2\pi i k}{h^\vee_G}\right),\qquad k=0,\ldots,h^\vee_G-1.
\end{equation}
where $h^\vee_G$ is the dual coxeter number of the gauge algebra.
A shift $\theta \to \theta+2\pi$ cyclically permutes the different vacua.
The low-energy superpotential is given by $W=c\Lambda^3$ where $c$ is a constant.
The AMSB deformation is then given by
\begin{equation}
    V_{\mathrm{tree}}=-3mW+\mathrm{c.c.}=-6cm|\Lambda|^3\cos(\theta/h^\vee_G).
\end{equation}
Thus, the energy of the $k$-th vacuum depends on $\theta$ and is given by the following:
\begin{equation}
    E_k(\theta)=-6cm|\Lambda|^3\cos\left(\dfrac{\theta+2\pi k}{h^\vee_G}\right).
\end{equation}
Consequently, the degeneracy among the $h^\vee_G$ vacua present in the SUSY limit is lifted and the vacua aquire distinct energy levels. The physical vacuum energy is therefore given by the minimum over branches, $\min E_k(\theta)$.
Therefore, as $\theta$ varies, branch crossing occurs.
At such a branch crossing, the vacuum energy itself remains continuous, but its derivative with respect to $\theta$ becomes discontinuous, signaling a first-order phase transition.
Note that in the SUSY limit, the gluino condensate spontaneously breaks the discrete $R$-symmetry as $\mathbb{Z}_{2h_G^\vee}\rightarrow\mathbb{Z}_2$ and consequently the $h_G^\vee$ vacua are degenerate.
On the other hand, once AMSB is introduced, the same breaking pattern is realized explicitly by the gluino mass.
Thus, there is only one vacuum in a generic point of $\theta$. There are two degenerate minima when $\theta = \pi$ mod $2\pi$.

\paragraph{AMSB in $\rSpin(k)$ with vectors}
Here we see the AMSB deformation of $\rSpin(k)$ theories with vector matters, focusing on the cases with $k-4$ and $k-3$ flavors \cite{Csaki:2021jax,Csaki:2021xuc}.
The supersymmetric dynamics of these theories were studied in \cite{intriligator1995duality}.
In the SUSY theory, there exists a moduli space parameterized by the meson fields
\begin{equation}
    M_{ij}=Q_iQ_j,\qquad i,j=1,\ldots,n_f ,
\end{equation}
which transform as a symmetric tensor of the flavor group $SU(n_f)$.
Then, the gaugino condensation generates the ADS superpotential
\begin{equation}
    W_{\rm ADS}=\pm\left(\frac{\Lambda^{3k-n_f-6}}{\det M}\right)^{\frac{1}{k-n_f-2}}.
\end{equation}
In the SUSY theory, they have a two-branch structure.
Each theory has a multibranch structure for a different reason.
For the theory with $k-4$ flavors, at a generic point on the moduli space, the gauge symmetry is broken to $\rSpin(4)\cong SU(2)\times SU(2)$.
Consequently, two ADS potentials are generated by gaugino condensation, and there is a branch on which they cancel each other exactly because of their opposite signs.
For the theory with $k-3$ flavors, the gauge symmetry is broken to $\rSpin(3)\cong SU(2)$.
In this case, the instanton contribution and the gaugino condensation contribution cancel each other exactly.

Turning on AMSB, the branch with an ADS superpotential is selected as the global minimum, and the resulting non-SUSY theory exhibits confinement together with chiral symmetry breaking.
More specifically, the vacuum is located at a meson expectation value proportional to the identity $M_{ij}\propto \delta_{ij}$.
As a consequence, the flavor symmetry is spontaneously broken as $SU(n_f)\longrightarrow SO(n_f)$.
The condensation of the fermion bilinear is encoded in the $F$-component of the meson superfield. 
Then, in the AMSB vacuum one finds $\langle \psi_i\psi_j\rangle\sim F_{M_{ij}}\propto\delta_{ij}$ which realizes the above symmetry breaking pattern.
Note that the minimum on this branch lies sufficiently far out on the moduli space that the K\"{a}hler potential can be approximated as canonical.
In this regime, the contribution arising from the superpotential in Eq.~\eqref{eq:AMSBtree} dominates, while corrections from non-canonical terms in the K\"{a}hler potential can be neglected.

On the $W=0$ branch, by contrast, the main AMSB effect comes from the non-canonical part of the K\"{a}hler potential.
See App.~\ref{sec:Kahler}. Without a handle on the analytic form of the K\"ahler potential we cannot judge whether the minimum is at the origin without symmetry breaking or away from the origin with spontaneous breaking of global symmetries. Yet we will see in many cases that it is highly implausible to have massless composite fermions $M_{ij}$ that match all anomalies when the constituent scalar acquires a mass in the UV. We suspect the minimum is away from the origin. 

Note that the vacuum energy in the $W=0$ branch with the potential due to non-canonical K\"ahler potential is $V \approx - m^2 \Lambda^2$. 
By contrast, the branch with the ADS superpotential contains deeper vacua whose vacuum energies are enhanced by a factor of $(\Lambda/m)^{2/(k-2)}$.
Hence, we regard the K\"{a}hler-potential contribution on the $W=0$ branch as subleading and conclude that this branch does not contain the true vacuum.

\section{$SU(6)$ with $A^{ijk}$}
\label{sec:SU(6)+A}

In this section, we consider $SU(6)$ theory with a rank-three antisymmetric tensor $A^{ijk}$.
Introducing AMSB, we study a resulting low-energy effective potential and a breaking pattern of a discrete symmetry.
We learn that such breaking pattern is different from SUSY theory studied in \cite{Dotti:1997wn,Csaki:1997cu,Henning:2021ctv,Ishikawa:2025yus}.
In this model, all the calculations can be carried out analytically, providing good practice for more three models in the following sections which is not solved analytically.

First of all, let us consider the SUSY theory.
Field contents are as following.
\footnote{
    More precisely, the discrete symmetry is $\bZ_3$ because the $\bZ_2$ subgroup of $\bZ_6$ is equivalent to the center.
}
\begin{equation}
\label{eq:charge_SU6}
\myspace{1.5}
\begin{tabular}{ScScScSc}
    symmetry & $SU(6)$ & $U(1)_R$ & $\bZ_6$ \\
    \hline
    $\mathcal{W}^\alpha$ & $\mathbf{adj}$ & $1$ & $0$ \\
    $A^{ijk}$ & $\mathbf{20}$ & $-1$ & $1$
\end{tabular}
\end{equation}
A classical flat direction can be chosen, by a $SU(6)$ gauge rotation, so that the only nonzero components are $u:=A^{123}$ and $v:=A^{456}$.
These expectation values break the gauge group as $SU(6)\rightarrow SU(3)\times SU(3)$.
Then, the number of moduli along this direction becomes $20-(35-8-8)=1$.
This fact implies the existence of a D-flat condition $|u|=|v|$.
Equivalently, there is one independent gauge-invariant modulus:
\begin{equation}
    A^4=c'\,u^2v^2,
\end{equation}
where $c'$ is a constant.
Note that there is no gauge invariant polynomial $A^2$ because the rank three antisymmetric representation of $SU(6)$ is pseudoreal.
The gaugino condensation of two $SU(3)$ factors generates ADS superpotentials leading nine branches.
Labeling such nine branches by $n_1,n_2=0,1,2$, the superpotential is given by
\begin{equation}
    W=(\omega^{n_1}-\omega^{n_2})\dfrac{\Lambda^{5}}{A^2},
\end{equation}
where $\omega$ is the cube roots of unity.
Note that the correct gauge invariant is $A^4$, and for convenience we denote $\sqrt{A^4}$ by $A^2$.
The theory has a discrete $\bZ_6$ symmetry.
However, by the low-energy dynamics, there exists a composite operator $\tr(A^2\mathcal{W}^\alpha\mathcal{W}_\alpha)$ which obtains vacuum expectation value \cite{Ishikawa:2025yus}.
Then, the discrete symmetry breaks to the $\bZ_2$ subgroup.
The anomaly matching conditions for continuous and the $\bZ_2$ symmetries are satisfied for the composite operator $A^4$.
Conversely, the failure of anomaly matching for the $\bZ_6/\bZ_2$ symmetry suggests that $\tr(A^2\mathcal{W}^\alpha\mathcal{W}_\alpha)$ condenses.
Note that this phenomena is seen in a $W=0$ branch which is realized in the case of $n_1=n_2$, while ADS branches exhibit runaway behavior.

Next, let us consider introducing AMSB.
For six branches which are ADS type, when introducing AMSB, we see that three vacua are generated and $\bZ_3$ symmetry is spontaneously broken.
From the first and third terms of Eq.~\eqref{eq:AMSBtree}, the AMSB contribution arising from the $F$-term is given by the following expression:
\begin{equation}
    V_{\text{AMSB}}=-5mW+\mathrm{h.c.}
\end{equation}
Using $u,v$ which are canonical in the K\"{a}hler potential, the total scalar potential becomes
\begin{equation}
    V=\dfrac{|c|^2|\Lambda|^{10}}{|u|^4|v|^2}+\dfrac{|c|^2|\Lambda|^{10}}{|u|^2|v|^4}+m\left(-5c\dfrac{\Lambda^5}{uv}\right)+\mathrm{h.c.}
\end{equation}
where $c=(\omega^{n_1}-\omega^{n_2})/\sqrt{c'}$ is a constant.
For the six possible choices of $n_{1,2}$ where $n_1\neq n_2$, the phase factor of $c$ becomes $e^{2\pi ik/12}$ where $k=1,3,\cdots,11$.
Defining $r,a$ as $u=re^{ia}$, the scalar potential as the function of $r,a$ is given by
\begin{equation}
\label{eq:V_SU6}
    V(r,a)=\dfrac{2|c|^2|\Lambda|^{10}}{r^6}-\dfrac{10|m||c||\Lambda|^5}{r^2}\mathrm{Re}[e^{i\mathrm{Arg}(m)}e^{2\pi ik/12}ie^{-2ia}e^{i\theta/3}].
\end{equation}
The minima are realized when $\mathrm{Re}[\cdots]=1$.
If so, the minimum of the potential can be found straightforwardly by differentiating it once with respect to $r$.
Then, it is given by
\begin{equation}
\label{eq:min_SU6}
    V_{\mathrm{min}}=-\dfrac{20\sqrt{15}}{9}\sqrt{cm^3\Lambda^5},\quad r_{\mathrm{min}}=\left(\dfrac{3|c||\Lambda|^5}{5|m|}\right)^{1/4}.
\end{equation}
When $m\ll\Lambda$, the minima are located far from the dynamical scale, and theory is expected to be weakly-coupled.
So it is almost sufficient to computate Eq.~\eqref{eq:AMSBtree} for the canonical K\"{a}hler potential.
In order to find solutions of $a_{\mathrm{min}}$ which minimize the potential, let us consider global symmetries and their action.
There are two global symmetries, that is, the fermion parity $\bZ_2^F$ which is a subgroup of the Lorentz group and $\bZ_6$ which rotates the angle of $A$.
Note that the AMSB breaks the $R$ symmetry to $\bZ_2^R$, however, this is not independent from $\bZ_6$ because it is realized by the combination of three times the generator of $\bZ_6$ and the fermion parity.
In addition, there is a $\bZ_2^W$ Weyl symmetry which is a subgroup of gauge symmetry swapping two $SU(3)$ factors.
These actions are given as following.
\begin{subequations}
\begin{alignat}{3}
    &\bZ_6&&:u\in(n_1,n_2)\longmapsto\zeta u&&\in(n_1+1,n_2+1) \\
    &\bZ_2^W&&:u\in(n_1,n_2)\longmapsto u&&\in(n_2,n_1)
\end{alignat}
\end{subequations}
where $\zeta=e^{2\pi i/6}$.
Note that $(n_1,n_2)$ distinguishes branches and $u$ is the moduli parameter which is a component of $A$.
Because the vacuum moduli related by gauge symmetry should be identified, the number of vacua is reduced by a factor of two due to the Weyl group.
In addition, the $\bZ_2$ subgroup of $\bZ_6$ is compensated by the center of the gauge group.
Note that we have omitted $\bZ_2^F$.
Since it is a subgroup of the Lorentz group, it is expected to remain unbroken.
In the following, we omit $\bZ_2^F$ when discussing spontaneous symmetry breaking.
Then, AMSB effect generates three vacua which are rotated by $\bZ_3$ in $\bZ_6$.
The symmetry breaking is given by
\begin{equation}
    \bZ_6\to\bZ_2.
\end{equation}
This fact agrees with Eq.~\eqref{eq:V_SU6}, since that formula essentially imposes the condition $2\pi k/12-2a\equiv0\mod2\pi$ and there are two solutions with fixed $k$ reproducing three vacua.

This symmetry breaking should be described in terms of a gauge invariant order parameter.
Indeed, using the vacuum expectation value of $u$ obtained above, we see that $A^4=c'u^2v^2$ acquires a nonzero value in the regime $m\ll\Lambda$ inducing $\bZ_6\to\bZ_2$.
This is also understood as a fermion composite operator $\tr({\psi_A}_{\alpha}{\psi_A}_{\beta}{\psi_A}^{\alpha}{\psi_A}^{\beta})$ where $\psi_A$ is the fermionic component of $A$.
Since $\psi_A$ survives in the non-SUSY limit, that composite operator can be tested in the lattice gauge theory.


On the other hand, on the branch where the superpotential vanishes, we need to implement AMSB using an action that includes a non-canonical K\"{a}hler potential. Since the theory has no symmetry left, anomaly matching condition does not constrain the massless spectrum. In principle the vacuum can be at the origin with a gapped spectrum, while the vacuum away from the origin is also possible depending on the form of the K\"ahler potential. In either case the spectrum is gapped without any symmetry.

\section{$Sp(k)$ with $A^{ij}$}
\label{sec:Sp(2k)+A}
In this section, we consider $Sp(k)$ gauge theories with a rank two antisymmetric tensor $A^{ij}$ \cite{Dotti:1997wn,Cho:1996bi,Csaki:1996eu}.
Here, we again employ AMSB to analyze the theory with broken SUSY.
Since an analytic treatment is difficult, we numerically determine the vacuum solutions for $k=3,4,5$.
We find that, once SUSY is broken, the symmetry breaking pattern differs from that obtained for the SUSY model in \cite{Ishikawa:2025yus}.

First of all, let us consider SUSY theory.
The classical moduli is parameterized by the $k-1$ gauge invariant polynomials $\tr(AJ)^i$ where $i=2,\cdots,k$.
Using the UV description, it is given by $a_i$ with a condition $\Sigma_ia_i=0$ where $\pm ia_i$ are eigenvalues of $A^{ij}$.
At a generic point of the moduli space, the gauge symmetry is spontaneously broken to $SU(2)^k$.
Thus, the complex dimension of the moduli space becomes $(k(2k-1)-1)-(k(2k+1)-3k)=k-1$.
Note that this number corresponds to the number of gauge invariant polynomials.
The charges of the matter are as following.\footnote{
    The precise discrete symmetry is $\bZ_{k-1}$ because the $\bZ_2$ subgroup of $\bZ_{2k-2}$ is in the center of the gauge group.
    However, unlike in Sec.~\ref{sec:SU(6)+A}, the number of vacua is not changed.
}
\begin{equation}
\label{eq:charge_Sp}
\myspace{1.5}
\begin{tabular}{ScScScSc}
    symmetry & $Sp(k)$ & $U(1)_R$ & $\bZ_{2k-2}$ \\
    \hline
    $\mathcal{W}^\alpha$ & $\mathbf{adj}$ & $1$ & $0$ \\
    $A^{ij}$ & $\ydiagram{1,1}$ & $-\dfrac{2}{k-1}$ & $1$
\end{tabular}
\end{equation}
The gaugino condensation of unbroken $SU(2)^k$ gauge symmetries generates ADS superpotential which is given by
\begin{equation}
    W=\sum_i\epsilon_i\dfrac{\Lambda^{k+2}}{\Pi_{j\neq i}(a_i-a_j)},
\end{equation}
where $\epsilon_i=\pm1$.
Depending on the choice of signs for $\epsilon_i$, there are $2^k$ branches.
When all the signs $\epsilon_i$ are identical, the superpotential vanishes exactly.
For other choice of the signs, the nonzero ADS superpotential is generated leading to runaway behavior.
This theory has a discrete $\bZ_{2k-2}$ symmetry.
Due to the strong dynamics in the IR, and one finds a condensing operator $\tr(A^{k-1}\cW^{\alpha}\cW_{\alpha})$ and the discrete symmetry $\bZ_{2k-2}$ is spontaneously broken to $\bZ_{k-1}$.
As a result, the anomaly matching conditions for $\tr(AJ)^i$ is satisfied for continuous symmetries, and not satisfied for the discrete $\bZ_{2k-2}/\bZ_{k-1}$ symmetry.

Next, we consider the AMSB deformation.
We learn that the symmetry breaking patterns are different from those of SUSY theories.
On branches other than the $W=0$ branch, the AMSB contribution from the $F$-term is obrtained from Eq.~\eqref{eq:AMSBtree} as
\begin{equation}
    V_{\mathrm{AMSB}}=m\left(\sum_la_l\dfrac{\partial W}{\partial a_l}-3W\right)+\mathrm{h.c.}
\end{equation}
Thus, the total scalar potential becomes
\begin{equation}
\label{eq:scalar_Sp}
\begin{split}
    V&=\sum_l\left|\sum_{i\neq l}\dfrac{\Lambda^{k+2}}{a_i-a_l}\left(\dfrac{\epsilon_i}{\Pi_{j\neq i}(a_i-a_j)}+\dfrac{\epsilon_l}{\Pi_{j\neq l}(a_l-a_j)}\right)\right|^2 \\
    &\quad-(k+2)m\sum_i\epsilon_i\dfrac{\Lambda^{k+2}}{\Pi_{j\neq i}(a_i-a_j)}+\mathrm{h.c.}
\end{split}
\end{equation}
Note that we omit the Lagrange multiplier associated with condition $\sum_ia_i=0$ for the irreducibility of $A^{ij}$.
Roughly, the minima of the potential can be estimated as follows
\begin{equation}
\label{eq:energy_SpkA}
    V_{\mathrm{min}}\sim(m^{2k}\Lambda^{2k+4})^{\frac{1}{k+1}},\quad a_{\mathrm{min}}\sim\left(\dfrac{\Lambda^{k+2}}{m}\right)^{\frac{1}{k+1}}.
\end{equation}
When $m\ll\Lambda$, the minima are on $a_{\mathrm{min}}\sim\Lambda(\Lambda/m)^{1/(k+1)}\gg\Lambda$.
Therefore, theory is weakly-coupled and the computation of $F$-term contribution is alomost sufficient because the K\"{a}hler potential is expected to be approximately canonical.
Now let us turn to symmetries.
There are one global symmetry except for $\bZ_2^F$ and one gauge symmetry.
The AMSB breaks $U(1)_R$ symmetry to $\bZ_{2k-2}^R$ for even $k$ and $\bZ_{k-1}^R$ for odd $k$.
However, this symmetry is realized as a combination of $\bZ_2^F$ and the phase rotation of the matter $\bZ_{2k-2}$, and is therefore not independent.
So, the true global symmetry is $\bZ_{2k-2}$.
Note that there is a remaining gauge symmetry $S_k^W$ which is a subgroup of the Weyl symmetry of $Sp(k)$.
We summarize the action of these discrete symmetries to the moduli space of vacua as following.
\begin{subequations}
\label{eq:action_Spk}
\begin{alignat}{3}
    &\bZ_{2k-2}&&:(a_1,\cdots,a_k)\in(\epsilon_1,\cdots,\epsilon_k)\longmapsto(\xi a_1,\cdots,\xi a_k)&&\in(-\epsilon_1,\cdots,-\epsilon_k) \\
    &S_k^W&&:(a_1,\cdots,a_k)\in(\epsilon_1,\cdots,\epsilon_k)\longmapsto(a_{\sigma(1)},\cdots,a_{\sigma(k)})&&\in(\epsilon_{\sigma(1)},\cdots,\epsilon_{\sigma(k)})
\end{alignat}
\end{subequations}
where $\xi=e^{2\pi i/2(k-1)}$, $(\epsilon_1,\cdots,\epsilon_k)$ means a branch with such a sign, and $\sigma(i)$ means an action of the symmetry group.
One might naively expect there to be $2k-2$ vacua related by $\bZ_{2k-2}$.
However, a more detailed analysis is actually required.
Since determining the minima of this scalar potential analytically is difficult, we instead determine them numerically for small values of $k$.
In particular, the global minima are obtained for $k=3,4,5$.
From this point onward, we discuss the three cases separately.

\begin{figure}[h]
    \centering
    \begin{tabular}{ccc}
        \includegraphics[width=0.3\linewidth]{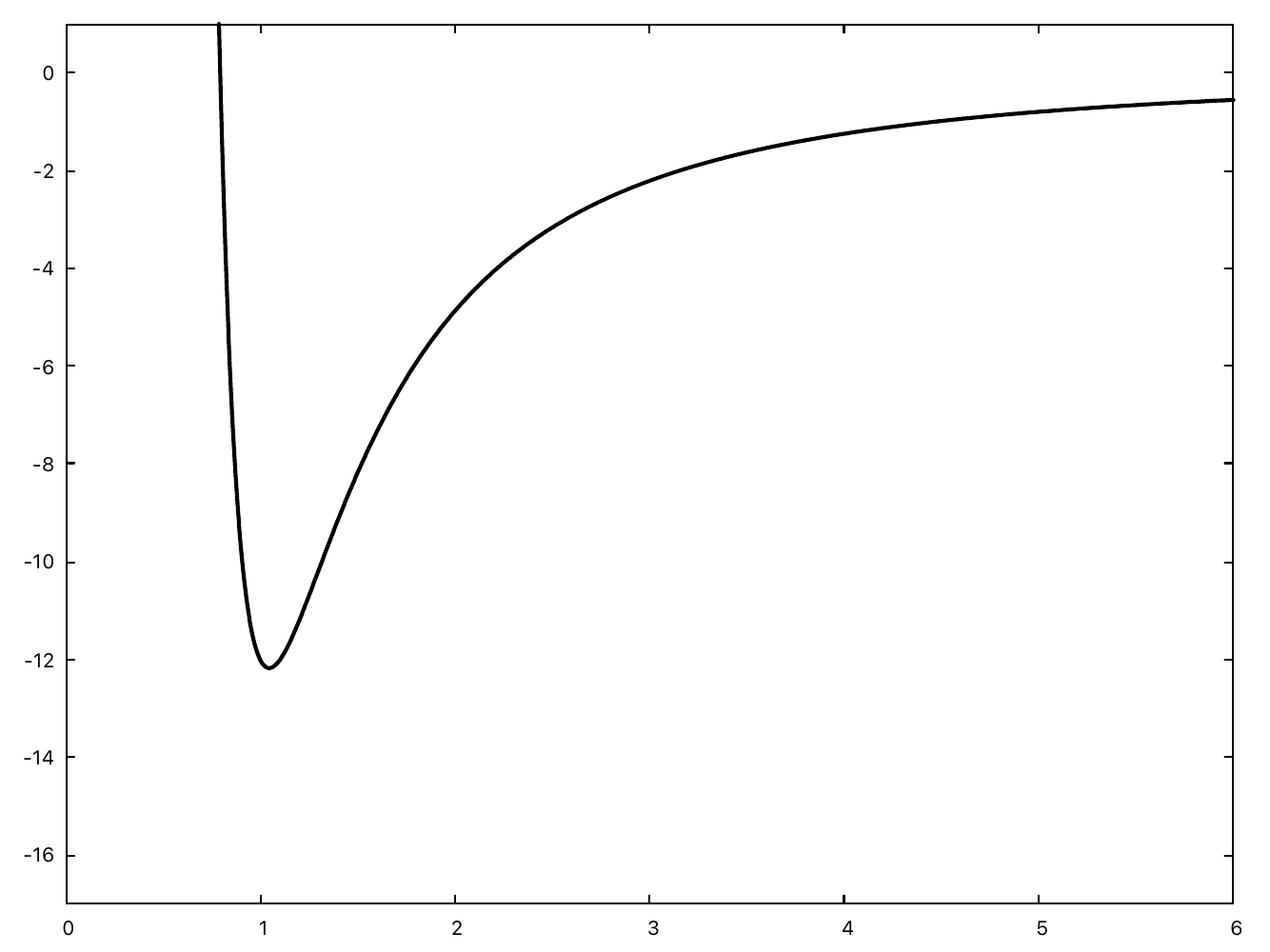}
        &
        \includegraphics[width=0.3\linewidth]{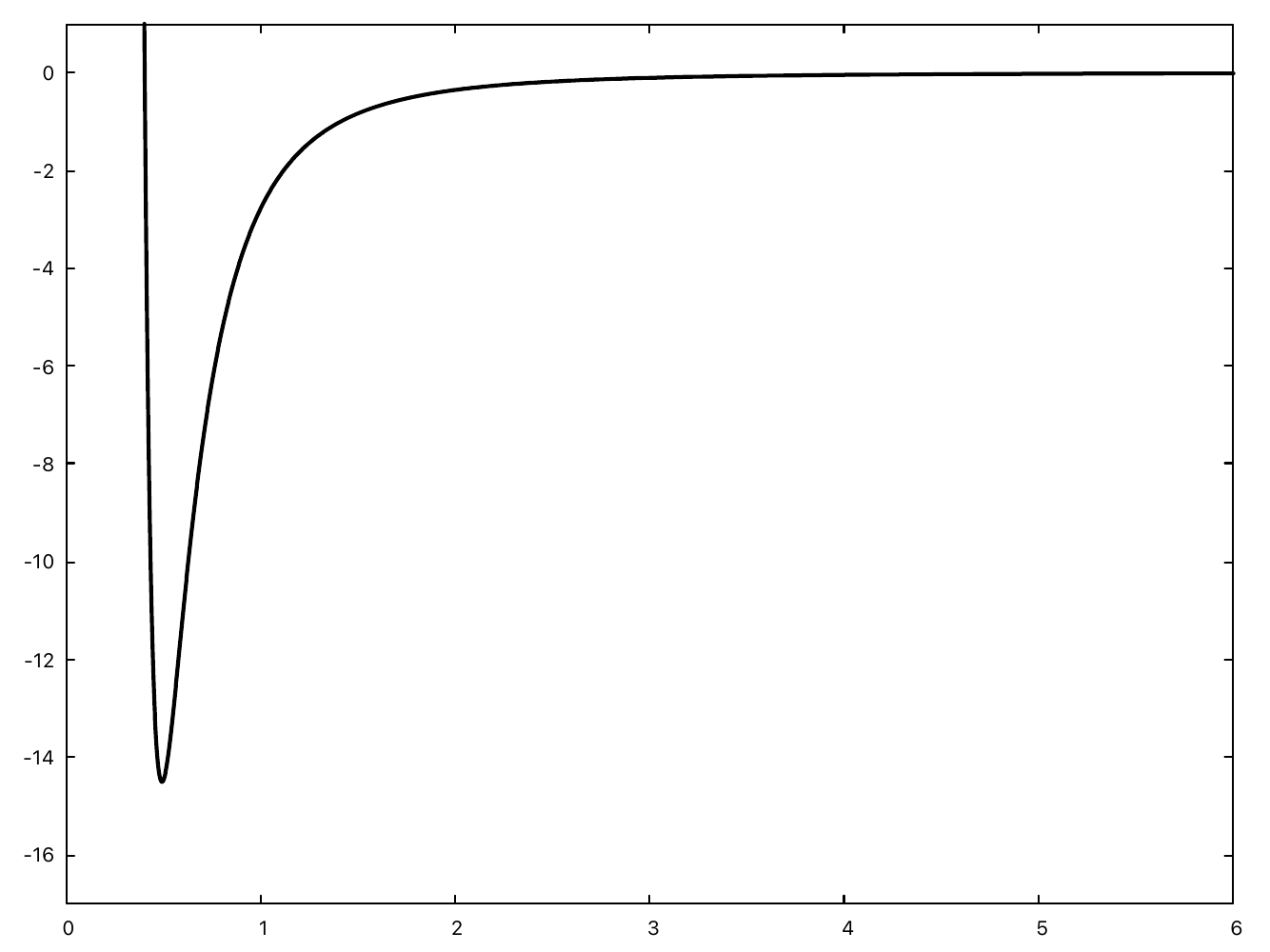}
        &
        \includegraphics[width=0.3\linewidth]{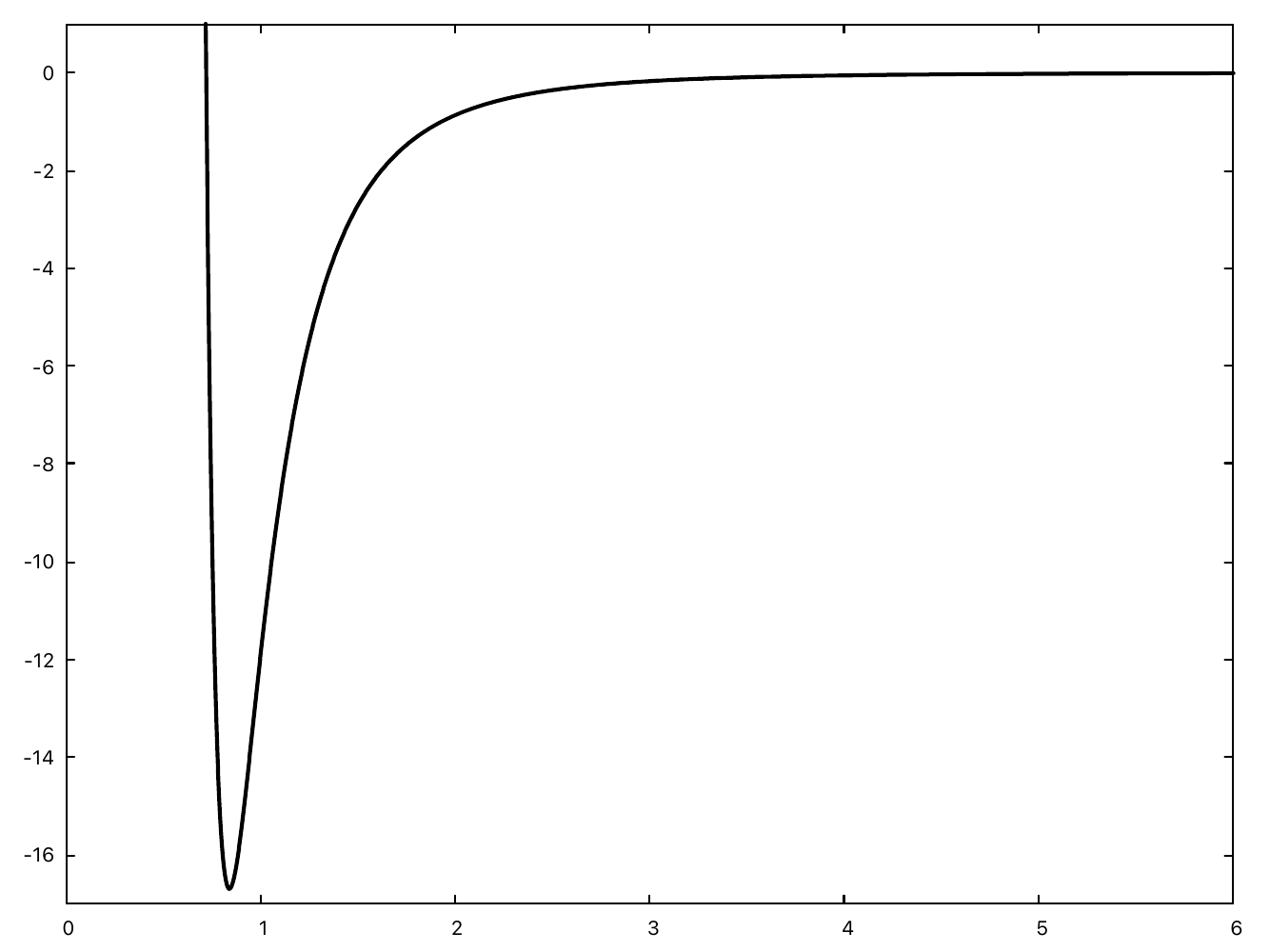}
        \\
        $Sp(3)+A$ & $Sp(4)+A$ & $Sp(5)+A$
    \end{tabular}
    \caption{The existence of vacua in $Sp(k)+A$ for $3\le k\le5$}
    \label{fig:spk}
\end{figure}

For $k=3$, the global minima are given by $V_{\mathrm{min}}\simeq-12.2(m^3\Lambda^5)^{1/2}$.
Note that there are $2^3=8$ branches with six exhibiting the ADS superpotential and two having no superpotential.
The minima are in ADS branches for the reasons explained later.
An example of values of the moduli parameters that realize the minima are given by
\begin{equation}
\begin{tabular}{c|ccc}
     & $a_1$ & $a_2$ & $a_3$ \\
    \hline
    $a_i/(\Lambda^5/m)^{1/4}$ & $1.0$ & $-1.0$ & $0.0$
\end{tabular}
\end{equation}
All the other minima are obtained by acting on this solution with the symmetry given in Eq.~\eqref{eq:action_Spk}.
The $\bZ_2\subset\bZ_4$ action on vacua are equivalent to a Weyl action which flips the sign of moduli within the same branch.
Therefore, the spontaneous breaking of the discrete symmetry is described by
\begin{equation}
    \bZ_4\to\bZ_2.
\end{equation}
The left panel of Fig.~\ref{fig:spk} shows the variation of the potential along a straight line extending from the origin to infinity which passes through one of the minima.

For $k=4$, the global minima are given by $V_{\mathrm{min}}\simeq-14.5(m^8\Lambda^{12})^{1/5}$.
Note that there are $2^4=16$ branches where two have vanishing superpotential, eight have ADS superpotential with $(\pm\pm\pm\mp)$ signs, and six have ADS superpotential with $(\pm\pm\mp\mp)$ signs.
There are local minima in the second and third ones, while the global minima are located in the third ones.
The moduli coordinates of the minima are, for example, given by
\begin{equation}
\begin{tabular}{c|cccc}
     & $a_1$ & $a_2$ & $a_3$ & $a_4$ \\
    \hline
    $a_i/(\Lambda^6/m)^{1/5}$ & $1.3$ & $-0.5$ & $0.5$ & $-1.3$
\end{tabular}
\end{equation}
All the other minima is obtained from the action Eq.~\eqref{eq:action_Spk}.
As seen in the $k=3$ case, the $\bZ_2$ subgroup of $\bZ_6$ is equivalent to an action of the Weyl group.
Therefore, the symmetry breaking pattern is given by
\begin{equation}
    \bZ_6\to\bZ_2.
\end{equation}
Thus, we obtain three physically equivalent vacua.
The center panel of Fig.~\ref{fig:spk} shows the behavior of the potential along a ray from the origin to infinity passing through one of the minima.
Finally, note that there are local minima in the branches with $(\pm\pm\pm\mp)$ signs, where its minima is given by $V_{\mathrm{min}}\simeq-12.6(m^8\Lambda^{12})^{1/5}$.

For $k=5$, the global minima are given by $V_{\mathrm{min}}\simeq-16.7(m^5\Lambda^7)^{1/3}$.
There are $2^5=32$ branches, where the superpotentials of two branches are zero.
There are local minima in the $10$ branches with $(\pm\pm\pm\pm\mp)$ signs.
In addition, the global minima are located in the $20$ branches with $(\pm\pm\pm\mp\mp)$ signs.
The global minima are realized at
\begin{equation}
\begin{tabular}{c|ccccc}
     & $a_1$ & $a_2$ & $a_3$ & $a_4$ & $a_5$ \\
    \hline
    $a_i/(\Lambda^7/m)^{1/6}$ & $1.5$ & $-1.5$ & $0.0$ & $0.8$ & $-0.8$
\end{tabular}
\end{equation}
and all vacua related to them by the symmetry transformation are also global minima.
The global symmetries are broken due to a similar mechanism to the case of $k=3$.
That is, the actions of $\bZ_2$ subgroups of $\bZ_8$ are equivalent to a Weyl action which flips the sign of moduli within the same branch.
Therefore, the discrete symmetries break as
\begin{equation}
    \bZ_8\to\bZ_2.
\end{equation}
Therefore, there are four vacua which are related by the action of $\bZ_4$.
The right panel of Fig.~\ref{fig:spk} shows the behavior of the potential along the direction passing through one of the minima.
Note that there are local minima in the branches with $(\pm\pm\pm\pm\mp)$ signs, and the minimum value is given by $V_{\mathrm{min}}\simeq-13.2(m^5\Lambda^7)^{1/3}$.

Based on the numerical analysis above, we can conjecture which branch contains the minimum for general $k$.
Let $n_+$ denote the number of signs $\epsilon_i$ equal to $+1$, and $n_-$ the number equal to $-1$.
We then find that all the minima identified in the analysis above lie on the branches for which $|n_+-n_-|$ is minimized.
This can be expected from the fact that the more signs $\epsilon_i$ coincide, the more polynomial terms in the superpotential cancel one another.
In particular, when all the signs are identical, the superpotential vanishes exactly.
Furthermore, there is also a pattern in the symmetry breaking structure.
Namely, the moduli parameters realizing the minima always appear in pairs with opposite signs.
When $k$ is odd, there is one additional zero.
As a result, the $\bZ_2$ subgroup of the $\bZ_{2k-2}$ symmetry is equivalent to an element of the Weyl group.
Thus, we obtain the following
\begin{equation}
    \bZ_{2k-2}\to\bZ_2.
\end{equation}
We therefore expect the broken symmetry to always be given by $\bZ_{k-1}$.
In the non-SUSY vacua, there are no fields that can match the anomaly, so the anomaly of the unbroken symmetry is expected to vanish.
We confirm that all anomalies involving the unbroken symmetries cancel.
\begin{equation}
\myspace{1.5}
\begin{tabular}{SlSl}
    anomaly & UV \\
    \hline
    $\bZ_2(\mathrm{gravity})^2$ & $1\cdot(k(2k-1)-1)\equiv0\,\mod1$ \\
    $\bZ_2^3$ & $1^3\cdot (k(2k-1)-1)\equiv0\,\mod1$ \\
\end{tabular}
\end{equation}
There is no requirement that the vacua contain any massless composite fermions.

This symmetry breaking should be characterized by a gauge invariant order parameter.
The condensation of an operator $\tr\left( AJ\right)^2$ is expected to induce the same pattern of symmetry breaking.
In fact, that operator condenses at least for $k=3,4,5$ cases.
As a fermionic operator, the operator $\tr\{(\psi_AJ)_{\alpha}(\psi_{A}J)^{\alpha}\}$ can be considered.
Note that this operator remains in the non-SUSY theory.

We finally discuss the contribution from the branches where the superpotential vanishes.
If the K\"ahler potential is such that the minimum is at the origin, the anomalies with the unbroken $\bZ_{k-1}$ symmetry must be matched. This was possible in the supersymmetric case with gauge-invariant polynomials ${\rm tr} (AJ)^{\ell}$ with $\ell=2, \cdots, k$. Obviously only the fermionic composites are needed for anomaly matching. Once AMSB is added, the scalar components of $A$
acquire positive mass-squared in the UV. We find it quite implausible that the composites ${\rm tr} (AJ)^{\ell}$ with scalar constituent can be massless. When $\ell$ is odd, the composite fermion can be made of fermionic constituent only and may be massless. Yet only with the subset of gauge-invariant polynomials anomalies do not match. Therefore we suspect the minimum is away from the origin with the same symmetry breaking pattern as the other branches. Recall that this branch does not give the global minimum anyway.

\section{$SU(2k)$ with $A^{ij}$ and $\tilde{A}_{ij}$}
\label{sec:SU(2k)+A+A}
In this section, we consider $SU(2k)$ theories with a rank two antisymmetric tensor $A^{ij}$ and its complex conjugate $\tilde{A}_{ij}$ \cite{Dotti:1997wn,Ishikawa:2025yus}.
We introduce AMSB and study the resulting low energy effective theories.
Again, we see that the discrete symmetries break in different way from their SUSY theories.
The discussions here are almost parallel to the previous section concentrating on $Sp(k)$ theories.
We analyze the effects of AMSB numerically for the cases of $k=3,4,5$.

Let us start with SUSY theories.
The moduli space is parameterized by the $k+1$ gauge invariant polynomials $\Pf A$, $\Pf\tilde{A}$, and $\tr(A\tilde{A})^i$ where $i=1,\cdots,k-1$.
In the UV description, it is parameterized by the eigenvalues of $A^{ij}$ and $\tilde{A}_{ij}$.
Note that both have the forms of the tensor product of a diagonal matrix and $\sigma_2$, because they are in antisymmetric representations of the gauge group.
We denote eigenvalues as $a_i$ and $\tilde{a}_i$ where $i=1,\cdots,k$.
There are $k$ D-flat conditions $|a_i|^2-|\tilde{a}_i|^2=c$ with a constant $c$.
At a generic point of the moduli space, the gauge symmetry is spontaneously broken to $SU(2)^k$.
The dimension of the moduli space is given by $2k(2k-1)-\{(4k^2-1)-3k\}=k+1$.
This agrees with the number of gauge invariant polynomials.
The charges of the matter are as following.
\footnote{
    Strictly, the discrete symmetry is $\bZ_{2k-2}$ due to the center, and reulting vacua becomes half as large.
}
\begin{equation}
\myspace{1.5}
\begin{tabular}{ScScScScSc}
    symmetry & $SU(2k)$ & $U(1)_R$ & $U(1)_V$ & $\bZ_{4k-4}$ \\
    \hline
    $\mathcal{W}^\alpha$ & $\mathbf{adj}$ & $1$ & $0$ & $0$ \\
    $A^{ij}$ & $\ydiagram{1,1}$ & $-\dfrac{1}{k-1}$ & $1$ & $1$ \\
    $\tilde{A}_{ij}$ & $\overline{\ydiagram{1,1}}$ & $-\dfrac{1}{k-1}$ & $-1$ & $1$
\end{tabular}
\end{equation}
Non perturbative strong dynamics in IR generate the following superpotential
\begin{equation}
    W=\sum_i\epsilon_i\dfrac{\Lambda^{2k+1}}{\Pi_{j\neq i}(a_i\tilde{a}_i-a_j\tilde{a}_j)},
\end{equation}
where $\epsilon_i=\pm1$.
Therefore there are $2^k$ branches similarly to the previous case.
When all the signs $\epsilon_i$ are identical, the superpotential vanishes, while all the other branches have the nonzero ADS superpotential.
In addition, on $W=0$ branch, there is a condensing operator $\tr\{(A\tilde{A})^{k-1}\cW^\alpha\cW_\alpha\}$ and the discrete symmetry $\bZ_{4k-4}$ is spontaneously broken to $\bZ_{2k-2}$.
As a result, the anomaly matching conditions for continuous and $\bZ_{2k-2}$ symmetries are satisfied for the composite operators $\Pf A$, $\Pf\tilde{A}$, and $\tr(A\tilde{A})^i$.

Let us consider the AMSB deformation.
Again, the symmetry breaking patterns are different from the SUSY theories.
On branches other than the $W=0$ branch, the AMSB contribution from the $F$-term is obtained from Eq.~\eqref{eq:AMSBtree} as
\begin{equation}
    V_{\mathrm{AMSB}}=m\left(\sum_la_l\dfrac{\partial W}{\partial a_l}+\sum_l\tilde{a}_l\dfrac{\partial W}{\partial\tilde{a}_l}-3W\right)+\mathrm{h.c.}
\end{equation}
Therefore, the total scalar potential is given by
\begin{equation}
\label{eq:scalar_SU}
\begin{split}
    V&=\sum_l(|a_l|^2+|\tilde{a}_l|^2)\left|\sum_{i\neq l}\dfrac{\Lambda^{2k+1}}{a_i\tilde{a}_i-a_l\tilde{a}_l}\left(\dfrac{\epsilon_i}{\Pi_{j\neq i}(a_i\tilde{a}_i-a_j\tilde{a}_j)}+\dfrac{\epsilon_l}{\Pi_{j\neq l}(a_l\tilde{a}_l-a_j\tilde{a}_j)}\right)\right|^2\\
    &\quad-(2k+1)m\sum_l\epsilon_l\dfrac{\Lambda^{2k+1}}{\Pi_{j\neq l}(a_l\tilde{a}_l-a_j\tilde{a}_j)}+\mathrm{h.c.}\\
    &=\sum_l(2|x_l|)\left|\sum_{i\neq l}\dfrac{\Lambda^{2k+1}}{x_i-x_l}\left(\dfrac{\epsilon_i}{\Pi_{j\neq i}(x_i-x_j)}+\dfrac{\epsilon_l}{\Pi_{j\neq l}(x_l-x_j)}\right)\right|^2\\
    &\quad-(2k+1)m\sum_l\epsilon_l\dfrac{\Lambda^{2k+1}}{\Pi_{j\neq l}(x_l-x_j)}+\mathrm{h.c.}
\end{split}
\end{equation}
where we define $x_i=a_i\tilde{a}_i$.
Note that the moduli space is spaned by $x_i$ and the constant $c$ which turns out to be zero in AMSB deformation.
Roughly speaking, both the locations and the value of the minima of the potential can be estimated by the following expressions
\begin{equation}
\label{eq:min_SUkAA}
    V_{\mathrm{min}}\sim(m^{2k-1}\Lambda^{2k+1})^{\frac{1}{k}},\quad a_{\mathrm{min}},\tilde{a}_{\mathrm{min}}\sim\left(\dfrac{\Lambda^{2k+1}}{m}\right)^{\frac{1}{2k}}.
\end{equation}
Note that when $m\ll\Lambda$, we ignore the contribution from the non-canonical terms of the K\"{a}hler potential.
Therefore, we only consider the third term of Eq.~\eqref{eq:AMSBtree} from now on.
The breaking of the discrete symmetries is different from the SUSY theory.
To see this, let us check the global symmetries and their action on antisymmetric fields.
Because we add AMSB terms, the $R$ symmetry breaks to $\bZ_{2k-2}^R$.
We also have the axial symmetry $\bZ_{4k-4}$, and the vector symmetry $U(1)_V$ which is expected not to break.
Again, the broken $R$ symmetry is not independent, and is realized by $\bZ_2^F$ and $\bZ_{4k-4}$.
In addition, there is a remaining gauge symmetry $S_k^W$ exchanging the eigenvalues.
The actions of these discrete symmetries on the moduli space are given as following.
\begin{subequations}
\label{eq:action_SUk}
\begin{alignat}{3}
    &\bZ_{4k-4}&&:(x_1,\cdots,x_k)\in(\epsilon_1,\cdots,\epsilon_k)\longmapsto(\xi x_1,\cdots,\xi x_k)&&\in(-\epsilon_1,\cdots,-\epsilon_k) \\
    &S_k^W&&:(x_1,\cdots,x_k)\in(\epsilon_1,\cdots,\epsilon_k)\longmapsto(x_{\sigma(1)},\cdots,x_{\sigma(k)})&&\in(\epsilon_{\sigma(1)},\cdots,\epsilon_{\sigma(k)})
\end{alignat}
\end{subequations}
where $\xi=e^{2\pi i/2(k-1)}$, $(\epsilon_1,\cdots,\epsilon_k)$ means a branch with such a sign, and $\sigma(i)$ means an action of the symmetry group.
The discussion is essencially the same for Sec.~\ref{sec:Sp(2k)+A}.
Note that the $\bZ_2$ subgroups of $\bZ_{4k-4}$ acts trivially to the moduli space without spontaneous breaking.
Thus, it is expected to be $2k-2$ vacua in this theory which turns out to be true.
In order to analyze the vacuum structure, we solve Eq.~\eqref{eq:scalar_SU} in a numerical way.
We compute the minima in the case of $k=3,4,5$.
We briefly summarize the results for each case.

\begin{figure}[h]
    \centering
    \begin{tabular}{ccc}
        \includegraphics[width=0.3\linewidth]{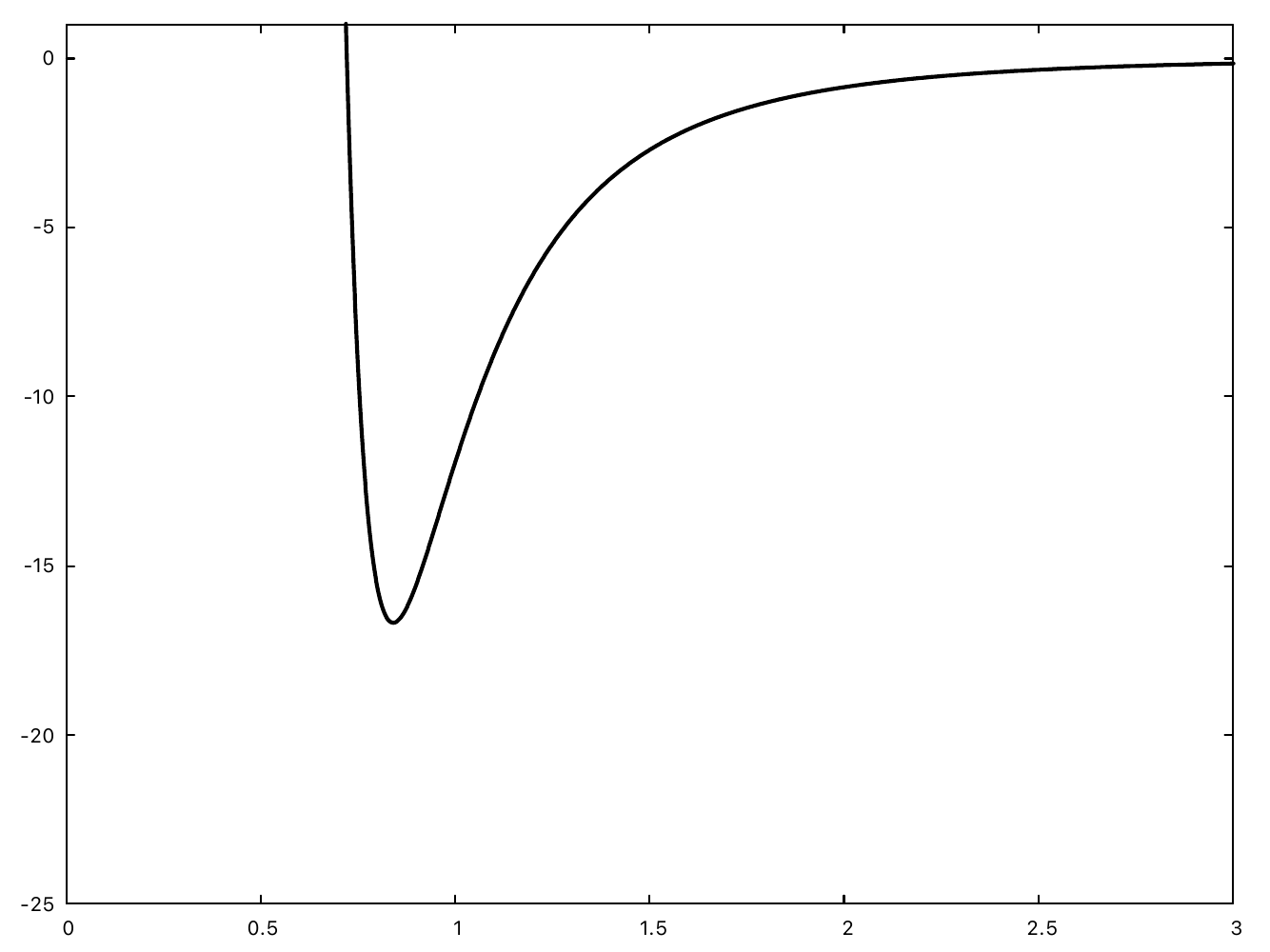}
        &
        \includegraphics[width=0.3\linewidth]{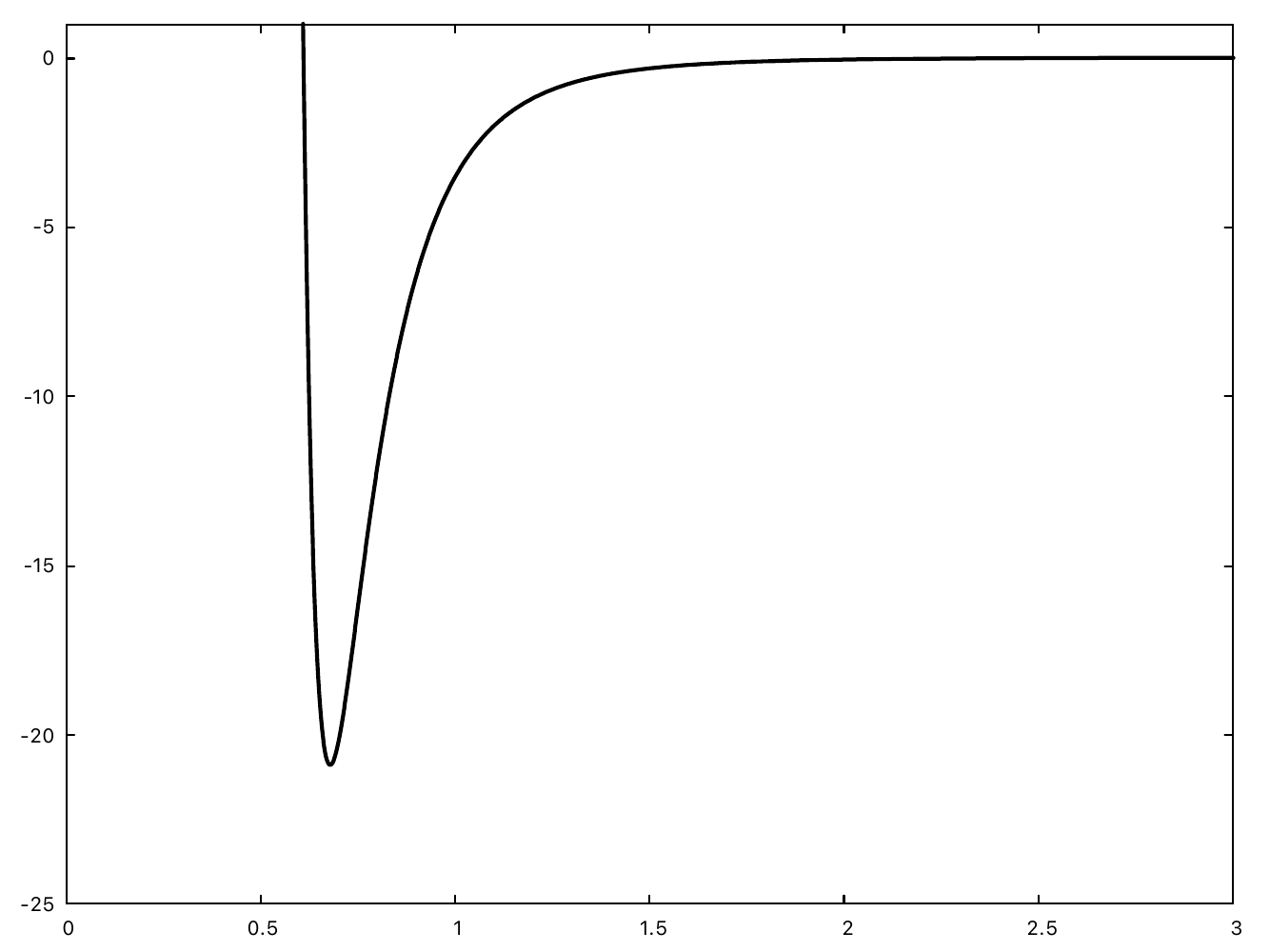}
        &
        \includegraphics[width=0.3\linewidth]{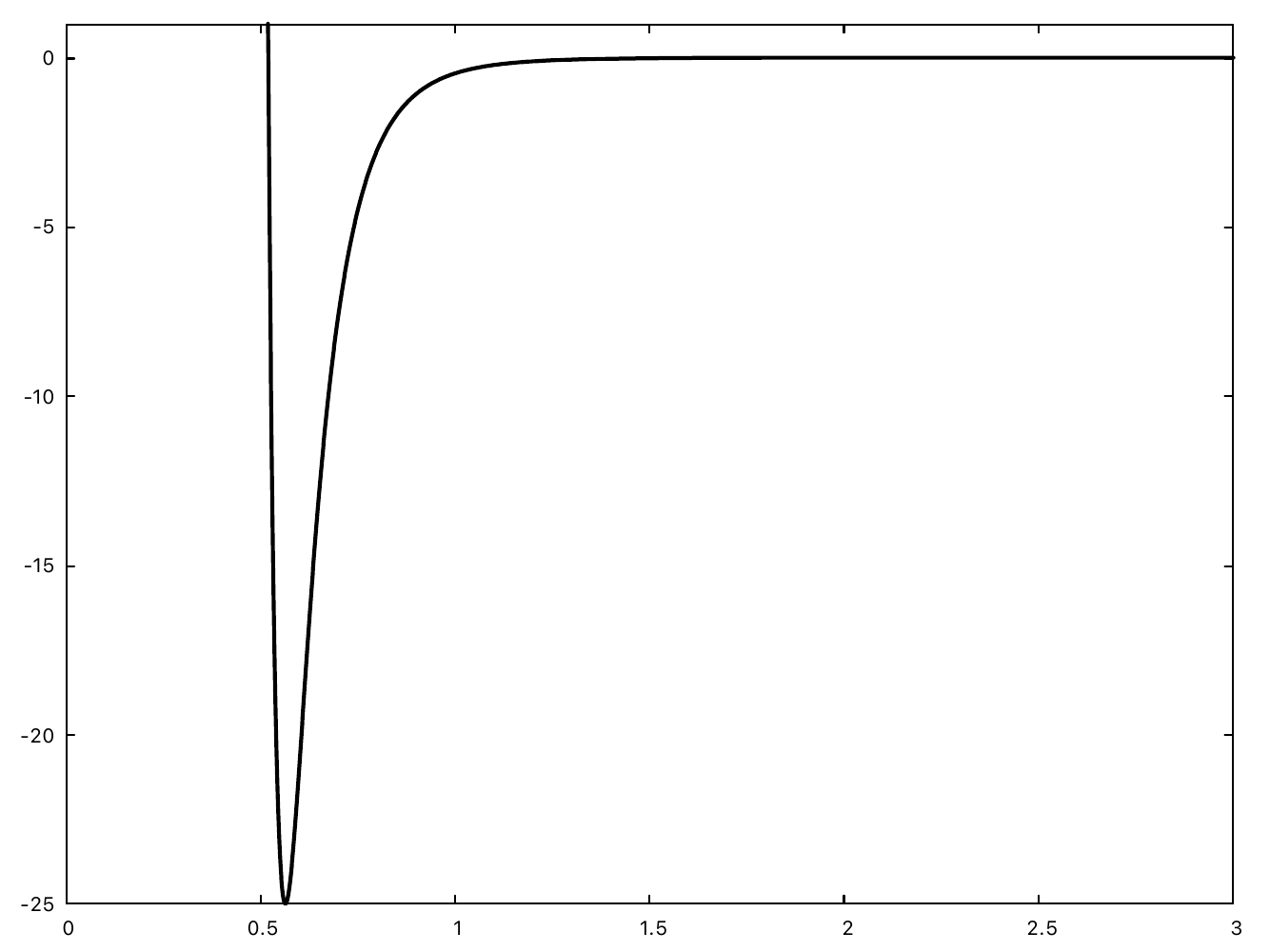}
        \\
        $SU(6)+A+\tilde{A}$ & $SU(8)+A+\tilde{A}$ & $SU(10)+A+\tilde{A}$
    \end{tabular}
    \caption{The existence of vacua in $SU(2k)+A+\tilde{A}$ for $3\le k\le5$}
    \label{fig:su2k}
\end{figure}

For $k=3$, the global minima are given by $V_{\mathrm{min}}\simeq-16.7(m^5\Lambda^7)^{1/3}$.
There are eight branches where six have ADS superpotential and two have vanishing superpotential.
The minima is in the ADS branches.
An example of values of the moduli parameters that realize the minima are given by
\begin{equation}
\begin{tabular}{c|ccc}
     & $x_1$ & $x_2$ & $x_3$ \\
    \hline
    $x_i/(\Lambda^7/m)^{1/3}$ & $1.2$ & $0.0$ & $1.2$
\end{tabular}
\end{equation}
All the other minima are obtained by the symmetry action Eq.~\eqref{eq:action_SUk}.
The broken symmetry $\bZ_4$ is given by the generator of $\bZ_8$, while the symmetry breaking is described as
\begin{equation}
    \bZ_8\to\bZ_2.
\end{equation}
The left panel of Fig.~\ref{fig:su2k} shows the potential profile along a ray passing through one of the minima.

For $k=4$, the global minima are given by $V_{\mathrm{min}}\simeq-20.9(m^7\Lambda^9)^{1/4}$.
This theory has 16 branches, where two are $W=0$ branches, eight are ADS branches, and the other six are ADS branches with global minima.
The moduli coordinates of the minima are given by
\begin{equation}
\begin{tabular}{c|cccc}
     & $x_1$ & $x_2$ & $x_3$ & $x_4$ \\
    \hline
    $x_i/(\Lambda^9/m)^{1/4}$ & $2.7$ & $0.5$ & $1.6$ & $0.0$
\end{tabular}
\end{equation}
All the other minima are obtained by acting the discrete symmetries to the above condidate.
The discrete symmetry spontaneously breaks as
\begin{equation}
    \bZ_{12}\to\bZ_2.
\end{equation}
Therefore, we have six vacua which are related by the action of the generator of the $\bZ_{12}$.
The center panel of Fig.~\ref{fig:su2k} shows the behavior of the potential along the direction passing through one of the minima.
Finally, note that there are local minima in $(\pm\pm\pm\mp)$ branches where the minimun value is given by $V_{\mathrm{min}}\simeq-19.4(m^7\Lambda^9)^{1/4}$.

For $k=5$, the global minima are given by $V_{\mathrm{min}}\simeq-25.0(m^9\Lambda^{11})^{1/5}$.
There are 32 branches, and three classes of branches which are not related by symmetries, that is, $W=0$ branches, ADS branches with local minima which have $(\pm\pm\pm\pm\mp)$ signs, and ADS branches with global minima which have $(\pm\pm\pm\mp\mp)$ signs.
The global minima are realized at
\begin{equation}
\begin{tabular}{c|ccccc}
     & $x_1$ & $x_2$ & $x_3$ & $x_4$ & $x_5$ \\
    \hline
    $x_i/(\Lambda^{11}/m)^{1/5}$ & $2.3$ & $0.4$ & $2.9$ & $1.2$ & $0.0$
\end{tabular}
\end{equation}
In addition, all the other global minima is related by the symmetry actions.
There are eight degenerate vacua, and the symmetry breaks as
\begin{equation}
    \bZ_{16}\to\bZ_2.
\end{equation}
The right panel of Fig.~\ref{fig:su2k} displays the potential profile along a direction passing through one of the minima.
Local minima also occur on the branches characterized by the sign choices $(\pm\pm\pm\pm\mp)$, with the minimum value computed as $V_{\mathrm{min}}\simeq-21.8(m^9\Lambda^{11})^{1/5}$.

From the above observations, we can formulate a conjecture as to which branch contains the minimum for general $k$.
All the minima found in our analysis so far lie on the branch for which $|n_+-n_-|$ is minimized.
This is quite similar to the result obtained in Sec.~\ref{sec:Sp(2k)+A}.
We can also conjecture the pattern of symmetry breaking for general $k$.
In contrast to the situation in the previous section, no regularity is observed in the moduli parameters.
So, it seems unlikely that any additional global symmetries become equivalent to elements of the Weyl group.
Therefore, we expect the symmetry breaking to be given by the naive estimate
\begin{equation}
    \bZ_{4k-4}\to\bZ_2.
\end{equation}
In addition, we check all anomalies of unbroken symmetries to be canceled.
This is consistent with the expectation that there are no massless fermions in the IR.
\begin{equation}
\myspace{1.5}
\begin{tabular}{SlSl}
    anomaly & UV \\
    \hline
    $U(1)_V(\mathrm{gravity})^2$ & $1\cdot k(2k-1)+(-1)\cdot k(2k-1)=0$\\
    $\bZ_2(\mathrm{gravity})^2$ & $2\cdot1\cdot k(2k-1)\equiv0\,\mod1$ \\
    $U(1)_V^3$ & $1^3\cdot k(2k-1)+(-1)^3\cdot k(2k-1)=0$\\
    $\bZ_2U(1)_V^2$ & $1\cdot1^2\cdot k(2k-1)+1\cdot(-1)^2\cdot k(2k-1)\equiv0\, \mod2$\\
    $\bZ_2^2U(1)_V$& $1^2\cdot1\cdot k(2k-1)+1^2\cdot(-1)\cdot k(2k-1)\equiv0\, \mod2$\\
    $\bZ_2^3$ & $2\cdot1^3\cdot k(2k-1)\equiv0\,\mod1$
\end{tabular}
\end{equation}

To describe above symmetry breaking in a gauge invariant operator, one candidate of a condensing operator is given by $\tr(A\tilde{A})$.
This condensation reproduces the same symmetry breaking.
Indeed, we find that $\tr(A\tilde{A})=\Sigma_i(a_i\tilde{a}_i)$ take nonzero values in the vacua for the case of $k=3,4,5$.
This symmetry breaking is also understood as a condensation of a fermion bilinear $\tr(\psi_{A}\psi_{\tilde A})$.
This operator survives in the non-SUSY limit, making it possible to trace its fate.

Although we have largely ignored it so far, it is important to take the K\"{a}hler contribution into account on the $W=0$ branch.
Just like with the $Sp(k)$ theories with $A^{ijk}$, we find it implausible that there is a full set of massless composite fermions to match all anomalies at the origin. We suspect the minimum is away from the origin with the same symmetry breaking pattern as the other branches. Recall that the $W=0$ branch does not produce a global minimum.

\section{$\rSpin(12)$ with $2S$}
\label{sec:Spin(12)+2s}
In this section, we consider $\rSpin(12)$ gauge theory with two matters in a spinor representation \cite{Dotti:1997wn}.
We introduce AMSB and study the resulting vacuum structure.
In this model, we analyze the effects of the AMSB deformation numerically, and again find the different symmetry breaking from the SUSY theory.

Let us start with SUSY theory.
There are two irreducible spinor representations of $\rSpin(12)$, corresponding to the two $\bZ_2$ factors in its center, that is, $\bZ_2\times\bZ_2$.
This theory contains two copies of one of these spinor representations.
That spinor matters tranform nontrivially under one $\bZ_2$ and trivially under the other $\bZ_2$.
Consequently, one unscreened $\bZ_2$ remains and the theory enters a $t$-confining phase.
The fields are given as following.
\footnote{
    Again, the precise global symmetry is $\bZ_8$ due to the center and the number of vacua is halved.
}
\begin{equation}
\label{eq:charge_SO12}
\myspace{1.5}
\begin{tabular}{ScScScScSc}
    symmetry & $\rSpin(12)$ & $SU(2)$ & $U(1)_R$ & $\bZ_{16}$ \\
    \hline
    $\mathcal{W}^\alpha$ & $\mathbf{adj}$ & $\mathbf{1}$ & $1$ & $0$ \\
    $S$ & $\mathbf{32}$ & $\mathbf{2}$ & $-\dfrac{1}{4}$ & $1$
\end{tabular}
\end{equation}
The moduli space is parameterized by the seven gauge invariant operators $S^2,S^4,S^6$, where $S^2$ and $S^6$ are $SU(2)$ singlet while $S^4$ is a quintet.
It is also parameterized by UV fields.
When considering one of the spinors, $\rSpin(12)$ is broken to $SU(6)$.
In addition, when we add one more spinor, the gauge symmetry becomes $SU(2)^3$.
Therefore, the complex dimension of the moduli space is computed by $32\cdot2-(66-3\cdot3)=7$ which corresponds to the number of the gauge invariant polynomials.
The classical moduli space is parameterized by the following eigenvalues
\begin{equation}
    S^\kappa=
    s^\kappa
    \;\oplus\;
    \begin{pmatrix}
        a^\kappa_1 && \\
        & a^\kappa_2 & \\
        && a^\kappa_3
    \end{pmatrix}
    \otimes\sigma_2
    \;\oplus\;
    \begin{pmatrix}
        \tilde{a}^\kappa_1 && \\
        & \tilde{a}^\kappa_2 & \\
        && \tilde{a}^\kappa_3
    \end{pmatrix}
    \otimes\sigma_2
    \;\oplus\;
    \tilde{s}^\kappa.
\end{equation}
where $\kappa=1,2$, and the eigenvalues are denoted as the representation of $SU(6)$.
From left to right, we consider $\mathbf{1}\oplus\mathbf{15}\oplus\overline{\mathbf{15}}\oplus\mathbf{1}$ ($=\mathbf{32}$).
The low-energy dynamics generates the effective superpotential which is given by
\begin{equation}
    W= \epsilon_1 \frac{\Lambda^{11}}{({\rm det} M_2)({\rm det} M_3)}
    + \epsilon_2 \frac{\Lambda^{11}}{({\rm det} M_3)({\rm det} M_1)}
    + \epsilon_3 \frac{\Lambda^{11}}{({\rm det} M_1)({\rm det} M_2)},
\end{equation}
where $\epsilon_i=\pm1$, and four dimensional matrices $M_i$ are defined as
\begin{equation}
    M_{1}= \left( \begin{array}{cccc}
        s^1 & \tilde{a}_1^1 & s^2 & \tilde{a}_1^2 \\ 
        \tilde{a}_2^1 & a_3^1 & \tilde{a}_2^2 & a_3^2 \\
        \tilde{a}_3^1 & a_2^1 & \tilde{a}_3^2 & a_2^2 \\
        a_1^1 & \tilde{s}^1 & a_1^2 & \tilde{s}^2 
        \end{array} \right), \quad
    M_{2}= \left( \begin{array}{cccc}
        s^1 & \tilde{a}_2^1 & s^2 & \tilde{a}_2^2 \\ 
        \tilde{a}_3^1 & a_1^1 & \tilde{a}_3^2 & a_1^2 \\
        \tilde{a}_1^1 & a_3^1 & \tilde{a}_1^2 & a_3^2 \\
        a_2^1 & \tilde{s}^1 & a_2^2 & \tilde{s}^2 
        \end{array} \right), \quad
    M_{3}= \left( \begin{array}{cccc}
        s^1 & \tilde{a}_3^1 & s^2 & \tilde{a}_3^2 \\ 
        \tilde{a}_1^1 & a_2^1 & \tilde{a}_1^2 & a_2^2 \\
        \tilde{a}_2^1 & a_1^1 & \tilde{a}_2^2 & a_1^2 \\
        a_3^1 & \tilde{s}^1 & a_3^2 & \tilde{s}^2 
        \end{array} \right).
\end{equation}
There are $2^3=8$ branches, and when all the signs $\epsilon_i$ are identical, the superpotential vanishes.
On this branch, one finds a condensing operator, $\tr(S^8\cW^{\alpha}\cW_{\alpha})$, whose nonzero vacuum expectation value breaks the discrete symmetry down to $\bZ_8$.
As a result, the anomaly matching conditions are satisfied for the $S^2,S^4,S^6$ operators.

\begin{figure}[h]
    \centering
    \includegraphics[width=0.3\linewidth]{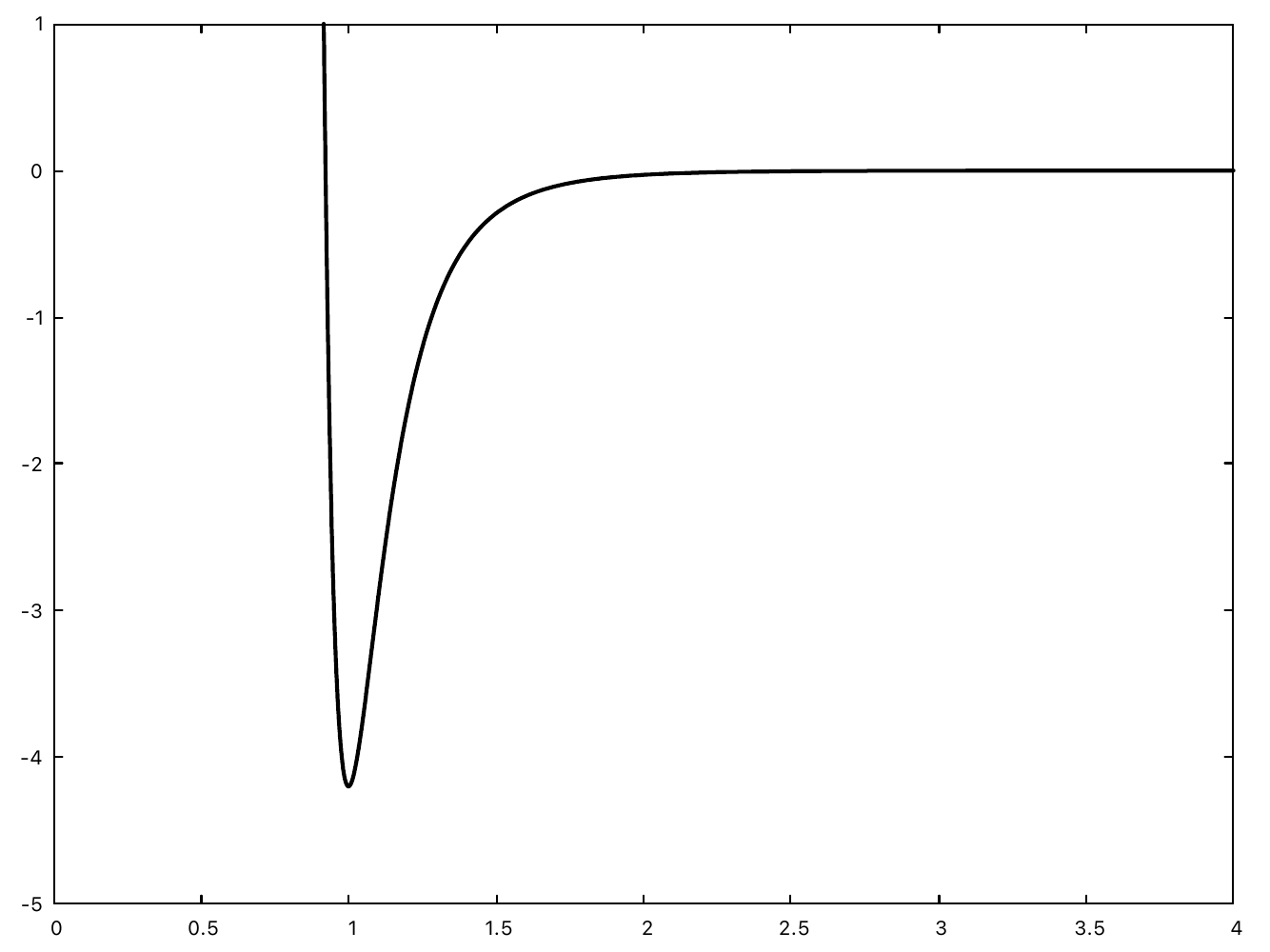}
    \caption{The existence of vacua in $\rSpin(12)+2S$}
    \label{fig:spin12}
\end{figure}

Next, let us consider the AMSB deformation.
As in the theories discussed above, the pattern of symmetry breaking is expected to differ between the SUSY and non-SUSY cases.
On branches with non-vanishing superpotential, the AMSB contribution from the $F$-term is given by
\begin{equation}
\begin{split}
    V_{\mathrm{AMSB}}=m\Big(&s^1\dfrac{\partial W}{\partial s^1}+\tilde{s}^1\dfrac{\partial W}{\partial \tilde{s}^1}+s^2\dfrac{\partial W}{\partial s^2}+\tilde{s}^2\dfrac{\partial W}{\partial \tilde{s}^2}\\
    &+\sum_{i=1}^{3}a_i^1\dfrac{\partial W}{\partial a_i^1}+\sum_{i=1}^{3}\tilde{a}_i^1\dfrac{\partial W}{\partial \tilde{a}_i^1}+\sum_{i=1}^{3}a_i^2\dfrac{\partial W}{\partial a_i^2}+\sum_{i=1}^{3}\tilde{a}_i^2\dfrac{\partial W}{\partial \tilde{a}_i^2}-3W\Big)+\mathrm{h.c.}
\end{split}
\end{equation}
Then, in order to find vacua, we need to analyze the minima of the resulting scalar potential.
In a rough analysis, the minima of the potential can be estimated as follows
\begin{equation}
    V_{\mathrm{min}}\sim(m^9\Lambda^{11})^{\frac{1}{5}},\quad a_{\mathrm{min}},\tilde{a}_{\mathrm{min}}\sim\left(\dfrac{\Lambda^{11}}{m}\right)^{\frac{1}{10}}.
\end{equation}
For $m\ll\Lambda$, the K\"{a}hler potential is expected to be approximately canonical, and the vacuum structure can be reliably determined by considering the $F$-term contribution alone.
Let us consider symmetries, which relates multiple vacua.
Due to the AMSB deformation, the $R$ symmetry becomes $\bZ_8^R\subset U(1)_R$.
This is not an independent symmetry to $\bZ_2^F$.
There are also the flavor $SU(2)$ symmetry which is broken to $SO(2)$, and the rotation of the chiral supermultiplet $\bZ_{16}$.
The Weyl group of $\rSpin(12)$ is $\bZ_2^5\rtimes S_6$, while its action to the moduli are given by $\bZ_2^3\rtimes S_3$ which is isomorphic to the full symmetry group of a cube.
\footnote{
    The action on the moduli, which will be discussed below, is explicitly given as follows.
    Let us consider a cube and label its eight vertices assigning $(+1,+1,+1)\to s$, $(-1,+1,+1)\to a_1$, $(+1,-1,+1)\to a_2$, $(+1,+1,-1)\to a_3$, $(+1,-1,-1)\to\tilde{a}_1$, $(-1,+1,-1)\to\tilde{a}_2$, $(-1,-1,+1)\to\tilde{a}_3$, $(-1,-1,-1)\to\tilde{s}$.
    Then, the $\bZ_2^3$ part correspods to the flipping of the sign of $x,y,z$ axes, while $S_3$ part correspods to the permutation of these axes.
    Mathematically, $\rtimes$ is what is known as a semidirect product, but no further knowledge of this structure is required in this paper.
}
The actions of these symmetries to the moduli are given as following.
\begin{subequations}
\label{eq:action_SUk}
\begin{alignat}{4}
    &\bZ_{16}&&:(s,\tilde{s},\cdots )\in(\epsilon_1,\epsilon_2,\epsilon_3)&&\longmapsto(\xi s,\xi\tilde{s},\cdots )&&\in(-\epsilon_1,-\epsilon_2,-\epsilon_3) \\
    &\bZ_2^3\rtimes S_3&&:(s,\tilde{s},\cdots )\in(\epsilon_1,\epsilon_2,\epsilon_3)&&\longmapsto(\Sigma(s),\Sigma(\tilde{s}),\cdots )&&\in(\epsilon_{\Sigma(1)},\epsilon_{\Sigma(2)},\epsilon_{\Sigma(3)}) \\
    &SU(2)&&:((s^1,s^2)^\top,\cdots )\in(\epsilon_1,\epsilon_2,\epsilon_3)&&\longmapsto(G(s^1,s^2)^\top,\cdots )&&\in(\epsilon_1,\epsilon_2,\epsilon_3)
\end{alignat}
\end{subequations}
where $\xi=e^{2\pi i/16}$, $G$ is a representation matrix of $SU(2)$, $\Sigma$ is an action of the Weyl group, and $(\epsilon_1,\epsilon_2,\epsilon_3)$ means a branch with such a sign.
The action of the nontrivial center of $SU(2)$ is equivalent to the $\bZ_2$ subgroup of $\bZ_{16}$.
To determine how these symmetries act on the minimum of the AMSB deformed moduli space, we numerically minimize the potential.
The global minima are given by $V_{\mathrm{min}}\simeq-4.2(m^9\Lambda^{11})^{1/5}$.
There are eight branches in total, six of which possess the ADS superpotential, while the remaining two have no superpotential.
The vacua are located on the ADS branches.
One representative set of moduli values realizing a minimum is given by
\begin{equation}
\begin{tabular}{c|ccccccc}
     & $\tilde{a}_1^1$ & $\tilde{a}_2^1$ & $\tilde{a}_3^1$ & $a_1^2$ & $a_2^2$ & $a_3^2$ & others \\
    \hline
    $s,\tilde{s},a,\tilde{a}/(\Lambda^{11}/m)^{1/10}$ & $1.9$ & $0.7$ & $1.2$ & $-1.9$ & $-0.7$ & $-1.2$ & $0.0$
\end{tabular}
\end{equation}
All the minima are related to one another by symmetry transformations.
In particular, ten of the sixteen moduli parameters vanish, while the remaining six nonzero values form three pairs.
Note that the vacuum is generated in such a way that the Cartan subgroup of the $SU(2)$ flavor symmetry remains unbroken, that is, the Cartan subgroup of the flavor $SU(2)$ symmetry is compensated by the broken gauge group.
Thus, the breaking of symmetries is given by
\begin{equation}
    \bZ_{16}\times SU(2)\to SO(2).
\end{equation}
The potential restricted to a line passing through a minimum is showed in Fig.~\ref{fig:spin12}.

We now verify that the anomalies are canceled among the UV fields.
The anomalies associated with the remaining global $SO(2)$ symmetry are summarized below.
As shown there, all of these anomalies are vanished suggesting that no massless composite fermions are required in the IR.
\begin{equation}
\myspace{1.5}
\begin{tabular}{SlSl}
    anomaly & UV \\
    \hline
    $SO(2)(\mathrm{gravity})^2$ & $32\cdot1+32\cdot(-1)=0$ \\
    $SO(2)^3$ & $32\cdot1^3+32\cdot(-1)^3=0$
\end{tabular}
\end{equation}

A gauge invariant description of the symmetry breaking should be obtained by identifying suitable order parameters.
The condensation of the operator $S^2$ induces the discrete symmetry breaking, but it does not induce the flavor symmetry breaking $SU(2)\to SO(2)$.
In contrast, the condensation of $S^4$ can break the $SU(2)$.
Therefore, the condensates of $S^2$ and $S^4$ realize the desired breaking.
Likewise, the condensation of $\psi_S^2$ and $\psi_S^4$ could reproduce the same symmetry breaking pattern.
Here, $\psi_S^2$ and $\psi_S^4$ is obtained by contracting the indices in the same way as for the scalar superfield.
Note that there may be other candidate composite operators that acquire vacuum expectation values.

On the other hand, on the branch where the superpotential vanishes, we must take into account the effects of AMSB arising from the K\"{a}hler potential. 
Similar to other cases, we find it implausible that we find massless composite fermions $S^2$, $S^4$, and $S^6$ with massive scalar constituents to match all anomalies. The minimum is likely to be away from the origin with the same symmetry breaking pattern as the other branches.  

\section{Discussion}
\label{sec:discussion}

In this paper, we studied AMSB in truly confining gauge theories.
In addition to reviewing pure Yang-Mills theory and $\rSpin(k)$ with vectors, we analyzed four classes of theories, that is, $SU(6)$ with $A^{ijk}$, $Sp(k)$ with $A^{ij}$, $SU(2k)$ with $A^{ij}$ and $\tilde{A}_{ij}$, and $\rSpin(12)$ with $2S$.
These theories provided controlled examples in non-SUSY $t$-confining theories.
Our analysis was performed in the regime $m\ll\Lambda$ where the exact SUSY description remains applicable.
A common feature of these theories was the existence of multiple branches. 
In the SUSY limit, there were broadly two types of branches in all theories, $W=0$ branches and ADS branches.
Once AMSB was introduced, the branches with ADS superpotential developed global minima.
We studied the vacuum structure in analytical and numerical ways, and studied patterns of symmetry breaking which is summarized in Table~\ref{$t$-confinement}.
\begin{table}[h]
    \centering
    \myspace{2}
    \begin{tabular}{ScSlSlSlSl}
        \hline
        & Theory & SUSY & With AMSB \\
        \hline
        (A) & $G$ & $\bZ_{2h}\to\bZ_2$ & no symmetries \\
        (B) & $\rSpin(k)+(k-4)V^i$ & \makecell[l]{$\bZ_{2k-8} \times SU(k-4) \times U(1)_R$ \\ $\to\bZ_{k-4} \times SU(k-4) \times U(1)_R$ } & \makecell[l]{$\bZ_{2k-8}\times SU(k-4)$\\$\to\bZ_{k-4}\times SO(k-4)$} \\
        (C) & $\rSpin(k)+(k-3)V^i$ & \makecell[l]{$\bZ_{2k-6} \times SU(k-3) \times U(1)_R$ \\ no breaking } & \makecell[l]{$\bZ_{2k-6}\times SU(k-3)$\\$\to\bZ_{k-3}\times SO(k-3)$} \\
        (D) & $SU(6)+A^{ijk}$ & $\bZ_6 \times U(1)_R\to\bZ_2 \times U(1)_R$ & $\bZ_6\to\bZ_2$ \\
        (E) & $Sp(k)+A^{ij}$ & $\bZ_{2k-2} \times U(1)_R\to\bZ_{k-1} \times U(1)_R$ & $\bZ_{2k-2}\to\bZ_2$ \\
        (F) & $SU(2k)+A^{ij}+\tilde{A}_{ij}$ & \makecell[l]{$\bZ_{4k-4}\times U(1)_V\times U(1)_R$\\$\to\bZ_{2k-2}\times U(1)_V\times U(1)_R$} & \makecell[l]{$\bZ_{4k-4}\times U(1)_V$\\$\to\bZ_2\times U(1)_V$} \\
        (G) & $\rSpin(12)+2S$ & \makecell[l]{$\bZ_{16}\times SU(2)\times U(1)_R$\\$\to\bZ_8\times SU(2)\times U(1)_R$} & $\bZ_{16}\times SU(2)\to SO(2)$ \\
        \hline
    \end{tabular}
    \caption{Summary of symmetry breaking patterns in SUSY and non-SUSY vacua}
    \label{$t$-confinement}
\end{table}
Theory (D) allowed for an analytic treatment and helped us understand the qualitative behavior of $t$-confining theories with AMSB.
For theories (E) and (F), we predicted that the vacua connected by symmetry transformations are given by $\bZ_{k-1}$ and $\bZ_{2k-2}$.
We also conjectured that the global minima occur on branches with nearly equal numbers of positive and negative signs in the ADS superpotential.
Since these theories had only one flavor, they did not possess a continuous moduli space.
On the other hand in theory (G), as in theories (B) and (C), the Nambu-Goldstone modes associated with the breaking of the flavor $SU(2)$ symmetry formed the vacuum moduli space.
These theories demonstrated that the symmetry breaking patterns due to AMSB were different from those of the SUSY theories. Namely a small but non-zero SUSY breaking abandons the SUSY vacua in the branch without superpotential and moves the vacua to the branch with an ADS-type superpotential.
This is a phase transition accompanied by symmetry breaking between the $m=0$ and $m\ll\Lambda$ vacua.

An important open question is whether the small AMSB regime studied here is continuously connected to the limit $m\gg\Lambda$ in which the superpartners decouple and the theories approach genuinely non-SUSY gauge theories.
If no phase transition occurs, our results give predictions for the symmetry breaking patterns of non-SUSY $t$-confining theories.

We regard our results to be candidate ground states for non-SUSY $t$-confining theories. We hope our work stimulates interest in the lattice gauge theory community to either confirm or refute the symmetry breaking patterns we presented here.

\acknowledgments

We thank Bea Noether who obtained the AMSB deformation of the theory (D) independently and confirmed our result. 
RI and SS are supported by Forefront Physics and Mathematics Program to Drive Transformation (FoPM), a World-leading Innovative Graduate Study (WINGS) Program, the University of Tokyo. RI is also supported by JST SPRING, Grant Number JPMJSP2108. SS is also supported by Research Fellow of Japan Society for the Promotion of Science (JSPS Research Fellow), JSPS KAKENHI Grant Number JP25KJ0857, and JSR Fellowship, the University of Tokyo. The work of H.M. was supported by the NSF grant PHY-2515115, by the U.S. Department of Energy (DE-AC02-05CH11231), by the JSPS Grant-in-Aid for Scientific Research JP23K03382, MEXT Grant-in-Aid for Transformative Research Areas (A) 26H00401, 26A204, 26H00403, Hamamatsu Photonics, K.K, Tokyo Dome Corporation, and by the World Premier International Research Center Initiative (WPI) MEXT, Japan.

\appendix

\section{K\"ahler Potential}
\label{sec:Kahler}

In the branch with no superpotential $W=0$, it may appear that there cannot be any potential even when AMSB is introduced because there are no interactions. This is not the case because the K\"{a}hler potential has an intrinsic scale $\Lambda$ that violates conformal invariance and hence leads to a non-trivial potential, see Eq.~\eqref{eq:AMSBtree}. It can produce a minimum at the origin or away from the origin depending on the functional form of the K\"ahler potential. If the minimum is at the origin, all anomalies must be matched with composite fermions. As discussed in the body of the paper, in many cases it is unlikely to have composite fermions once supersymmetry is broken, and hence a minimum away from the origin is preferred. 

Suppose a moduli space with no superpotential parameterized by a gauge-invariant polynomial $X$, which is a composite of $n$ elementary chiral superfields. If it is part of the degrees of freedom that match anomalies, its K\"ahler potential must be non-singular at the origin, namely
\begin{align}
    K \approx c X^* X,
\end{align}
where $c\sim O(1)$ in the unit where $\Lambda = 1$. On the other hand, $X$ describes a $D$-flat direction and its K\"ahler potential must behave as 
\begin{align}
    K \approx c' (X^* X)^{1/n}
\end{align}
when $X \gg 1$. We cannot calculate the behavior of $K$ when $X \sim O(1)$, while we know it must be positive definite
\begin{align}
    K_{XX^*} = \frac{\partial^2 K}{\partial X \partial X^*} > 0.
\end{align}

A simple interpolation between the two limiting regimes is
\begin{align}
    K = f(\xi) = n c (1+ a^n\xi)^{1/n}, \qquad \xi = X^* x
\end{align}
where $n c a = c'$. The AMSB induces a potential
\begin{align}
    V &= m^2 \left( K_i g^{i\bar{\jmath}} K_{\bar{\jmath}} - K\right). 
\end{align}
It is easy to see that the potential is very flat at the origin, $V \propto V_0 + \lambda (X^*X)^2 + O(X^* X)^3$. Therefore the actual shape is very sensitive to the detailed form of the interpolating function. 

In fact, a small modification
\begin{align}
    K = f(\xi) = n c (1 + b \xi + a^{2n}\xi^2)^{1/2n}
    \label{eq:Kahler}
\end{align}
gives a positive definite metric while a minimum away from the origin (see plots). 

\begin{figure}[t]
\centerline{
\includegraphics[width=0.45\textwidth]{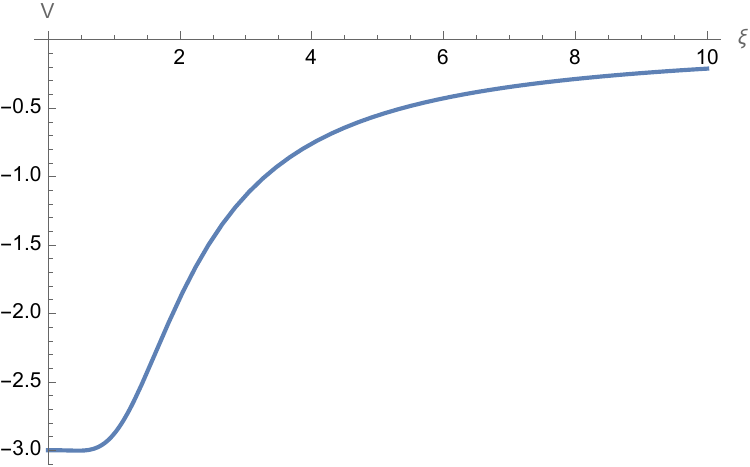}
\includegraphics[width=0.45\textwidth]{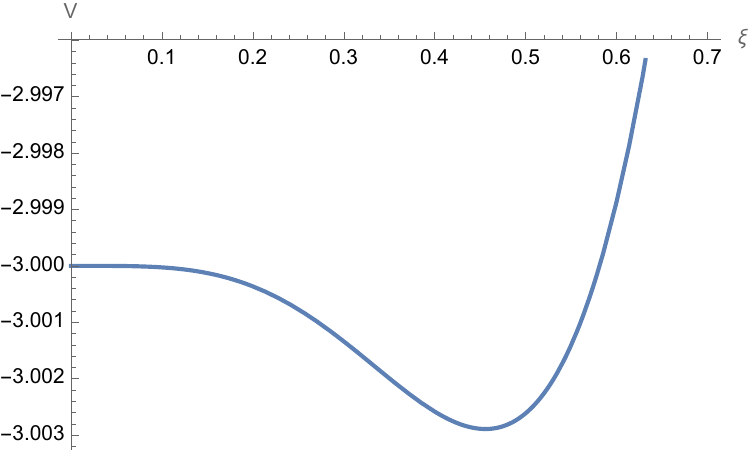}
}
\caption{The AMSB potential $V$ as a function $\xi=X^* X$ without a superpotential with the K\"ahler potential Eq.~\eqref{eq:Kahler} with $c=a=b=1$, $n=3$ in the unit $m=1$ (normalization of the potential) and $\Lambda=1$ (normalization of the field $X$). The left plot shows the very flat potential near the origin while the potential asymptotes to zero from below at the inifininty. The right plot shows a minimum away from the origin due to $b\neq 0$.}
\label{fig:Kahler}
\end{figure}

\bibliography{./bibliography}
\bibliographystyle{JHEP}

\end{document}